\documentclass[10pt,twocolumn,letterpaper]{article}

\usepackage[pagenumbers]{cvpr} 

\usepackage{graphicx}
\usepackage{amsmath}
\usepackage{amssymb}
\usepackage{booktabs}
\usepackage{multirow}
\usepackage{tabularx}
\usepackage{tabulary}
\usepackage{pifont}
\usepackage[toc,page]{appendix}
\usepackage{xspace}

\newcommand{\cmark}{\ding{51}}%
\newcommand{\xmark}{\ding{55}}%

\newcommand*\rot{\rotatebox{65}}
\newcommand*\textvert{\rotatebox{90}}

\def\name{Neuralizer\xspace}
\DeclareMathOperator{\resunit}{ResUnit}
\DeclareMathOperator{\concat}{cat}

\binoppenalty=\maxdimen
\relpenalty=\maxdimen

\usepackage[pagebackref,breaklinks,colorlinks]{hyperref}

\usepackage[capitalize]{cleveref}
\crefname{section}{Sec.}{Secs.}
\Crefname{section}{Section}{Sections}
\Crefname{table}{Table}{Tables}
\crefname{table}{Tab.}{Tabs.}

\def\cvprPaperID{9629} 
\def\confName{CVPR}
\def\confYear{2022}

\begin{document}

\title{\name: General Neuroimage Analysis without Re-Training}

\author{Steffen Czolbe\\
University of Copenhagen and MGH\\
{\tt\small per.sc@di.ku.dk}
\and
Adrian V. Dalca\\
MIT, and MGH, Harvard Medical School\\
{\tt\small adalca@mit.edu}
}
\maketitle
\global\csname @topnum\endcsname 0
\global\csname @botnum\endcsname 0

\begin{abstract}
   Neuroimage processing tasks like segmentation, reconstruction, and registration are central to the study of neuroscience. Robust deep learning strategies and architectures used to solve these tasks are often similar. Yet, when presented with a new task or a dataset with different visual characteristics, practitioners most often need to train a new model, or fine-tune an existing one. This is a time-consuming process that poses a substantial barrier for the thousands of neuroscientists and clinical researchers who often lack the resources or machine-learning expertise to train deep learning models. In practice, this leads to a lack of adoption of deep learning, and neuroscience tools being dominated by classical frameworks.
   
   We introduce \name, a single model that generalizes to previously unseen neuroimaging tasks and modalities without the need for re-training or fine-tuning. Tasks do not have to be known a priori, and generalization happens in a single forward pass during inference. The model can solve processing tasks across multiple image modalities, acquisition methods, and datasets, and generalize to tasks and modalities it has not been trained on. Our experiments on coronal slices show that when few annotated subjects are available, our multi-task network outperforms task-specific baselines without training on the task.
\end{abstract}

\begin{figure}[t]
  \centering
  \includegraphics[width=1\linewidth]{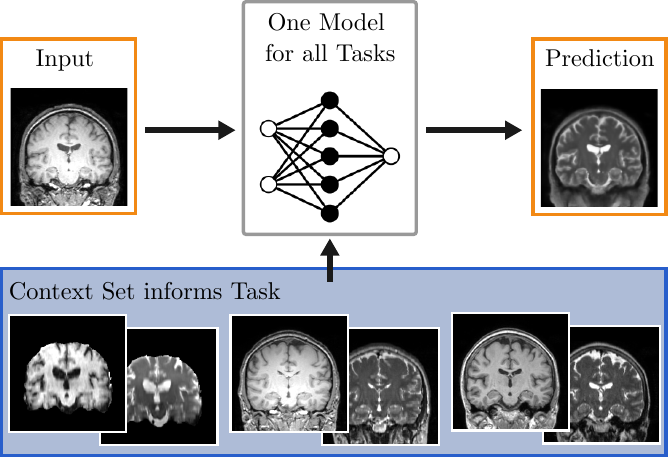}
  \caption{\name can solve a broad range of image processing tasks, including new ones not seen during training, with a single model by conditioning the prediction on a context set of examples. After training on a diverse set of tasks, the model can generalize to new tasks in a single forward pass without re-training or fine-tuning. The model is highly flexible, requiring no prior definition of the set of tasks, and can be conditioned with context sets of any length.}
  \label{fig:abstract_method}
\end{figure}

\section{Introduction} 
Computational methods for the processing and analysis of neuroimages have enabled a deep understanding of the human brain. The field has also led to advanced patient care by facilitating non-invasive methods of diagnosis and treatment. Recent deep learning research promises to substantially increase the accuracy and speed of neuroimaging analysis methods.

\begin{figure*}[t]
  \centering
   \includegraphics[width=1\textwidth]{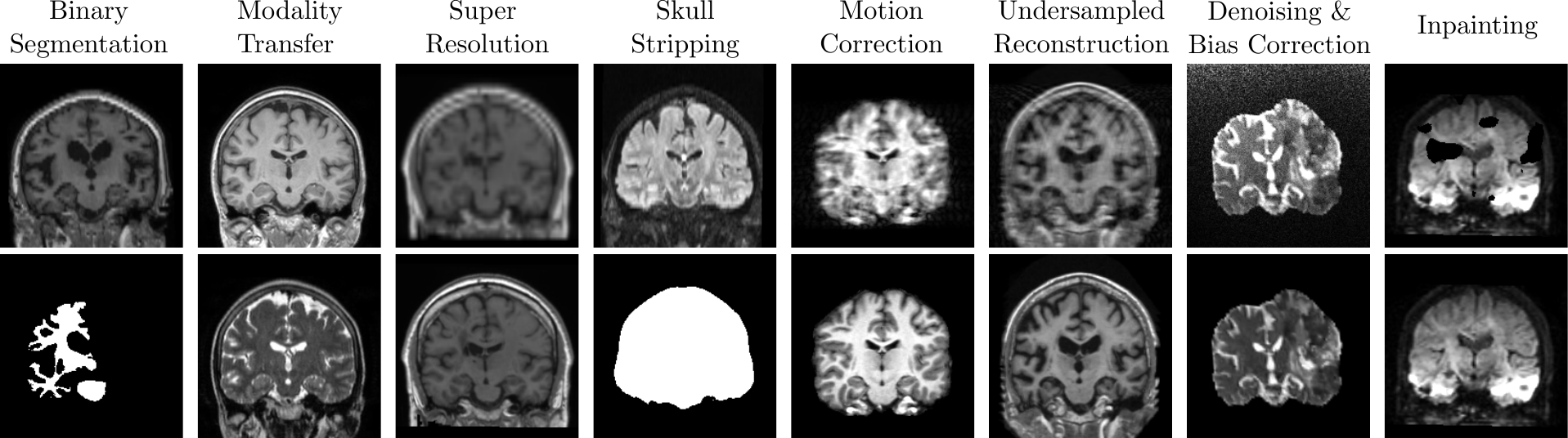}
   \caption{Example neuroimaging tasks and modalities included in our dataset (top: input images, bottom: output images).}
   \label{fig:task_examples}
\end{figure*}

A drawback of most current deep-learning-based approaches is that each model is limited to solving the task it has been trained on, on the type of data it has been trained on. Generalization to new tasks and domains, such as different acquisition protocols or new segmentation, is a main barrier to adoption~\cite{Li2020a}. Performing neuroimaging tasks like segmentation, registration, reconstruction, or motion correction requires different models for each processing step, despite operating on the same input data and methods exhibiting strong similarities in network architecture~\cite{Ronneberger2015,Hoffmann2020a, Billot2021}. Yet, designing and training models to solve these tasks on each dataset is prohibitively expensive. To train a deep learning model, a dataset needs to be compiled and often manually annotated, and the network, training, and data loading logic needs to be implemented. All these steps generally require machine learning and neuroimaging expertise. In addition, computational resources like specialized graphics processing hardware and software infrastructure needs to be available. These requirements are particularly problematic in clinical research settings due to a high cost of annotation and  a lack of machine learning expertise and hardware. The many closely related neuroimaging tasks and image modalities and acquisition characteristics require custom solutions, many of which are not available. 
As a consequence, many works forgo using methods adapted to their task and data characteristics, and instead use existing methods even when their data acquisition falls outside of the protocols used for building the tool \cite{Glasser2016,VanErp2015,balakrishnan2019voxel}. 
As neuroimaging tasks have much in common, generalization is a promising proposal to reduce the number of models that have to be trained.

\subsection*{Contribution}
We introduce \name, a general-purpose neuroimaging model that can solve a broad range of neuroimaging tasks on diverse image modalities (\cref{fig:task_examples}), without the need for task-specific training or fine-tuning. \name can solve new tasks, unseen during training, using a set of examples of the new task at inference (\cref{fig:abstract_method})

\name involves a convolutional architecture (\cref{fig:model}), that takes as input a context set of examples that define the processing task, and thus does not require prior specification of the tasks. The method enables single-pass generalization during inference and can process any number of reference images in a single pass to inform the prediction.

As a first method tackling task generalization in neuroimaging, we focus on analyzing the capabilities of such system and presenting general insights, and limit our experiments to 2D. We evaluate our model by comparing the single-pass generalization performance to task-specific baselines conditioned on an equivalent amount of data. We find that \name outperforms the baselines on tasks where $\leq 32$ labeled examples are available, despite never training on the task. When generalizing to new segmentation protocols, \name matches the performance of baselines trained directly on the dataset.

\section{Related Work}
We give a short introduction to neuroimaging tasks, terminology, and methods. We then provide an overview of fundamental methods for adapting a model to multiple domains, including multi-task learning, few-shot learning, fine-tuning, and data synthesis. 

\subsection{Neuroimage analysis}
Neuroimage analysis employs computational techniques to study the structure and function of the human brain. Common imaging techniques are structural magnetic resonance imaging (MRI), functional MRI, diffusion tensor imaging (DTI), computed tomography (CT), and Positron emission tomography (PET). Each imaging method can create diverse images with different characteristics and contrasts, which are further diversified depending on the properties of the acquisition site~\cite{Lee2021,Yuan2018}, device, protocol, imaging sequence~\cite{Kuijf2019}, and use of contrast agents~\cite{Gong2018,Brats2018}.

To analyze these images, a variety of processing tasks are most often combined in a processing pipeline. Common processing tasks include anatomical segmentation~\cite{fischl2012freesurfer,tzourio2002automated,pohl2007hierarchical,billot2020learning,dalca2018anatomical,dalca2019unsupervisedseg,Chen2018,wachinger2018deepnat}, skull stripping~\cite{Hoopes2022,Segonne2004,Smith2002,Kleesiek2016,Hoopes2022,Yogananda2019}, defacing~\cite{Hale2011,Abramian2019}, registration~\cite{Klein2009,faisalBeg2005,hoopes2022learning,Arsigny2006,Ashburner2007,Avants2008,balakrishnan2019voxel,Hoffmann2020a,Czolbe2021b}, modality transfer~\cite{Nie2017,Sun2019,Osman2022}, in-painting~\cite{Liu2021,Nguyen2021,Liu2015,Han2018,Yang2016}, super-resolution~\cite{Manjon2010,Manjon2010a,dalca2018medical,Wang2022,Laguna2022}, reconstruction, and de-noising~\cite{Lustig2008,Singh2022,Laguna2022}, bias field removal~\cite{Learned-Miller2004a,Goldfryd2021}, surface fitting~\cite{Hoopes2022a} and parcellation~\cite{Shen2013,Tzourio-Mazoyer2002}.

Multiple toolboxes provide a suite of interoperable software components, most implementing classical optimization strategies. Widely used toolboxes include Freesurfer~\cite{fischl2012freesurfer}, FSL~\cite{Jenkinson2012,Smith2004,Woolrich2009}, SPM~\cite{Frackowiak2004}, CIVET~\cite{Ad-Dabbagh2006}, BrainSuite~\cite{Shattuck2002}, HCP pipeline~\cite{VanEssen2013}, and BrainIAK~\cite{Kumar2021}. Deep-learning-based methods are starting to be included because of their improved accuracy and shorter runtime~\cite{Billot2021,Hoopes2022}. While these methods provide solutions for common neuroimaging applications, most are limited to a single task and few modalities. Developers need to manually update the pipelines to include new processing tasks and to support a wider variety of image modalities. This process requires extensive technical expertiese and computational resources, often not available to the clinical neuroscientists focusing on scientific questions.

\subsection{Multi-task learning}
Multi-Task Learning (MTL) frameworks solve multiple tasks simultaneously by exploiting similarities between related tasks~\cite{Caruana1997}. MTL can improve performance and reduce computational cost and development time compared to designing task-specific solutions~\cite{Evgeniou2004,Sener2018}. In neuroimaging, MTL networks were recently proposed for the simultaneous segmentation and classification of brain tumors by training a single network with separate prediction heads associated with the different tasks~\cite{Gupta2021,Diaz-Pernas2021}. This strategy is challenging to scale as the number of tasks increases, requires prior determination of the set of tasks, and importantly does not enable generalization of the model to new tasks. With \name, we build on these methods to achieve scalable MTL, without the need for multiple network heads, and importantly with the ability to generalize to new tasks and modalities.

\subsection{Fine-tuning}
To tackle problems in the limited data scenarios frequent in medical imaging, neural networks can be pre-trained on a related task with high data availability and then fine-tuned for specific tasks. For example, a common approach involves taking a Res-Net~\cite{He2016} trained on ImageNet~\cite{deng2009imagenet} and fine-tuning part of the network for a new task~\cite{Huh2016,Kornblith2019,Vinyals2016a}. For medical imaging, networks pre-trained on large sets of medical images are available~\cite{Chen2019}, and fine-tuning them to new tasks results in shortened training time and higher accuracy~\cite{Alzubaidi2021,Mahajan2020}. However, fine-tuning also requires machine learning expertise and computational resources, most often not available in clinical research. Additionally, in scenarios with small datasets, fine-tuning models trained on large vision datasets can be harmful~\cite{Raghu2019}.

\subsection{Few-shot learning}
Few-shot models generate predictions from just a few labeled examples~\cite{Ravi2017,Wang2020,Liu2019a,Schonfeld2019}, or in the case of zero-shot methods~\cite{Bucher2019}, none at all. Many of these methods require training or fine-tuning. In computer vision, several methods pass a query image, along with a set of support images and labels as input to the model~\cite{Liu2019a,Snell2017,Seo2022,Vinyals2016}. Natural image segmentation methods~\cite{Liu2020,Zhang2019} use single image-label pairs~\cite{Zhang2019a,Li2021} as support or aggregate information from a larger support set~\cite{Li2020}. Recent few-shot learning methods in the medical image segmentation setting~\cite{Bian2022,Feyjie2020,Feng2021} operate on a specific anatomical region in a single image modality~\cite{Zheng2017,Hansen2022}. 
Similar Prior-Data Fitted Networks (PFNs) are fitted to multiple datasets at once to learn the training and prediction algorithm~\cite{Muller2022}. During training, this strategy draws a dataset, a set of data points and their labels from it, masks one of the labels and predicts it. The resulting model aims to generalize to new datasets. PFNs have only been applied to low-dimensional and tabular data~\cite{Hollmann2022}.
Our solution builds on ideas from these methods, but aims to solve a much larger range of diverse image-to-image tasks on neuroimages of many modalities.

\subsection{AutoML methods}
AutoML tools can be used to automate the steps of implementation, training, and tuning deep learning models, reducing the technical knowledge required of the user. \mbox{NN-UNet}~\cite{Isensee2020a} automates the design and training of models for biomedical image segmentation, and has been successfully applied to brain segmentation~\cite{Isensee2021,El-Hariri2022,Luu2022}. While AutoML effectively reduces the technical requirements for the implementation, massively parallel hardware is still required for performing the internal hyper-parameter search and training the model. Additionally, AutoML methods reduce the flexibility in solution design, as they are often specific to a type of task or data structure.

\subsection{Data augmentation and synthesis}
Data augmentation increases the diversity of training data by augmenting or modifying existing data~\cite{Ronneberger2015,ZhaoMIT2019}. It improves model robustness to input variability that may not be available in the original training data. In neuroimaging, arbitrary image modalities can be simulated by synthesis of images without requiring any real data~\cite{Billot2021,Hoopes2022,Campello2020,Skandarani2020,Hoffmann2020a}. In meta-learning, data augmentation can further be used to generate entirely new tasks~\cite{Liu2020a,Yao2021,universeg2022}. We use data augmentations and further expand existing methods by developing rich neuroimaging task augmentations for generalization to unseen neuroimaging tasks.

\begin{figure*}[t]
  \centering
    \includegraphics[width=1\linewidth]{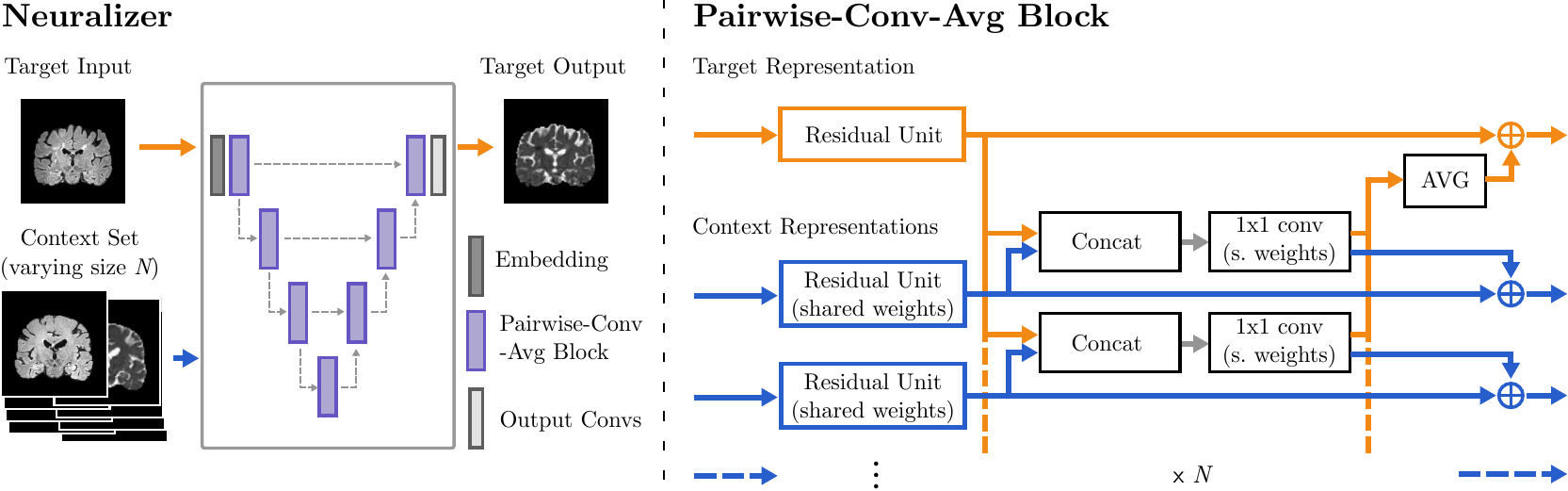}

   \caption{\name consists of 7 Pairwise-Conv-Avg blocks (right), arranged in a U-Net-like~\cite{Ronneberger2015,Milletari2016} configuration (left). Each Pairwise-Conv-Avg block enables interaction between the input image and the image pairs present in the context set. The block consists of a residual unit, pairwise convolution of each context member with the target, and an averaging of results across the context set to update the representation. The architecture is invariant to context size $N$.}
   \label{fig:model}
\end{figure*}

\section{\name}
We introduce \name, a multi-task model for neuroimage analysis tasks. In this section, we first define the training framework and adaptations necessary to operate on a diverse range of tasks and input types. We then introduce the model architecture, training, and inference strategies.

\subsection{Generalizabe multi-task model}
Let $T$ represent a set of tasks, with a subset of tasks $T_{\text{seen}}$ seen during training. Each task consists of input-output image pairs $(x_t, y_t)$ from potentially multiple underlying datasets with input and output spaces $x_t \in \mathcal{X}, y_t \in \mathcal{Y}$.

To enable generalization to unseen tasks, we condition the model on a context set $C_t = \{(x_{t,i}, y_{t,i})\}_{i=1}^{N}$ of input-output image pairs passed to the model alongside the prediction task. The context set defines the desired task, and can vary in size $|C_t| = N$ and is re-sampled from the underlying task-datasets for each input. \cref{fig:abstract_method} gives an example for a modality transfer task.

We employ a neural network $g_{\theta}(x_t, C_t) = y_t$ with weights $\theta$ that applies the task defined by context set $C_t$ to the input neuroimage $x_t$. We optimize the network using supervised training with the loss
\begin{equation}\label{eq:objective}
    \mathcal{L}(T_{\text{seen}}; \theta) = 
    \mathbb{E}_{t \in T_{\text{seen}}} \big[ \mathbb{E}_{(x_t, y_t, C_t)} [ \mathcal{L}_t(y_t, g_{\theta}(x_t, C_t))] \big],
\end{equation}
where $\mathcal{L}_t$ is a task-specific loss function.

\subsection{Design for diverse tasks}
To process different tasks with a single model, we carefully select the loss function, neuroimage encodings, and generation of the training set for each task type.

\vspace{-0.2cm} \paragraph{Loss function.} \label{sec:loss_funs}
\name solves both segmentation tasks (e.g. anatomical segmentation and skull-stripping via a brain mask), more general and image-to-image tasks (e.g. denoising). We use the Soft Dice Loss~\cite{Milletari2016} for segmentation-like tasks, and the pixel-wise Mean Squared Error $MSE(y_t, g(x_t, C_t)) = \frac{1}{2 \sigma^2} \sum_p [{y_t}_p - {g(x_t, C_t)}_p]^2$ with balancing hyperparameter $\sigma^2$ for other tasks. As the network optimizes multiple tasks during training, the balance of the loss terms can dramatically affect the optimization and resulting performance. 

\vspace{-0.2cm}  \paragraph{Input and output encoding.}
For \name to work on both segmentation and image-to-image tasks, we facilitate simultaneous input of multiple image modalities and masks. We design the input space $\mathcal{X}$ to accept floating point value images with three channels, and zero-pad any channels unnecessary for a specific task. The output space $\mathcal{Y}$ follows the same design but uses only one channel.


\vspace{-0.2cm}  \paragraph{Training dataset creation.}
At each training iteration, we first sample a task $t$ from $T_{\text{seen}}$, selecting the task-specific dataset (\cref{tab:training-data}). From this dataset, we sample the input image, ground truth output, and image pairs for the context set. To increase the range of images that can be used to condition the trained model, the image modalities and acquisition protocols of entries of the context set can differ from the input image for some tasks. Supplemental section~\ref{supplement:train_data} contains a detailed description of the training data generator.

\begin{table}[]
\caption{Tasks, Modalities, Datasets, and Segmentation classes used in this paper, and involved in training \name.}
\label{tab:training-data}
\begin{tabularx}{\linewidth}{lX}
\cmidrule(r){1-1}\cmidrule{2-2}
Tasks                                & Modalities   \\ \cmidrule(r){1-1}\cmidrule{2-2}
Binary Segmentation              & T1-w.        \\
Modality Transfer                    & T2-w.        \\
Super Resolution                      & MRA          \\
Skull Stripping                      & PD           \\
Motion Correction      & FLAIR        \\
Undersampled Reconstruction         & ADC          \\
Denoising \& Bias correction                            & DWI          \\
Inpainting                           & DTI (17 dir.)\\ \cmidrule(r){1-1}\cmidrule{2-2}
&\\
\end{tabularx}

\begin{tabularx}{\linewidth}{lX}
\cmidrule(r){1-1}\cmidrule{2-2}
Datasets                                & Segmentation Classes   \\ \cmidrule(r){1-1}\cmidrule{2-2}
OASIS 3~\cite{Marcus2007,hoopes2022learning}      & \multirow{2}{4.75cm}{Freesufer protocol, 31 classes~\cite{Billot2021, fischl2012freesurfer}}                       \\
BRATS~\cite{Brats2015,Brats2017,Brats2018}        &                                                                                          \\
IXI~\cite{IXI}          &    \\
ATLAS R2.0~\cite{Liew2022}    &  \multirow{2}{4.75cm}{Manually-annotated Hammers Atlas, 96 classes~\cite{Hammers2003,Gousias2008,Faillenot2017}} \\
Hammers Atlas\cite{Hammers2003} &                                                                                          \\
WMH Challenge~\cite{Kuijf2019} &                                                                                        \\
ISLES2022~\cite{Petzsche2022}    &       Brainmasks~\cite{Hoopes2022, fischl2012freesurfer}   \\ \cmidrule(r){1-1}\cmidrule{2-2}
\end{tabularx}
\end{table}

\subsection{Model architecture}
\cref{fig:model} shows the \name network architecture, adapted with the concurrently developed~\cite{universeg2022} -- a method that focuses on solving broad segmentation tasks. As the architecture is independent of the task, we omit the task subscript in this section.

The input image $x$ and the image pairs of the context set $C_i=(x_i, y_i), i=1,...,N$ are first passed through an embedding layer consisting of a single $1 \times 1$ convolution with learnable kernels $e_x, e_C$, to obtain the representations $r_{x} = x \ast e_x$, $r_{C_i} = \concat(x_i, y_i)  \ast e_C$ where $\ast$ is the convolution operator. This combines each context image pair to a joint representation $r_{C_i}$ and maps all representations to a uniform channel width $c$, which is constant throughout the model. Next, we process the representations using multiple Pairwise-Conv-Avg Blocks (explained below), arranged as a U-Net-like configuration~\cite{Ronneberger2015,Milletari2016} to exploit multiple scales. The output $r^{{\text{out}}}_{x}$ of the final Pairwise-Conv-Avg Block is processed by a residual unit~\cite{He2016} and a final $1 \times 1$ conv layer to map to one output channel. All residual units consist of two $3 \times 3$ conv layers, a shortcut connection, and GELU activation functions~\cite{Hendrycks2016}. 

Compared to standard CNNs, \name uses a mechanism to enable knowledge transfer from the context set to the input image. We design the Pairwise-Conv-Avg Block (\cref{fig:model}, right) to model this interaction. The block maps from representations of the target input $r^{\text{in}}_{x}$ and context pairs $r^{\text{in}}_{C_i}$ to output representations $r^{\text{out}}_{x}$, $r^{\text{out}}_{C_i}$ of the same size. First, we process each input separately with a residual unit to obtain $r^{\text{int}}_{x} = \resunit_x(r^{\text{in}}_{x})$ and $r^{\text{int}}_{C_i} = \resunit_C(r^{\text{in}}_{C_i})$. The residual units, which involve two convolutions, operate on the context representations and have shared weights. Second, we pairwise concatenate the context representations with the target representation on the channel dimension: $p_i = \concat(r^{\text{int}}_{x}, r^{\text{int}}_{C_i})$.  We combine the pairwise representations and reduce the channel size back to $c$ using a $1 \times 1$ convolution with learnable kernel $k_x$, and update the target representation by averaging across context members $r^{\text{out}}_{x} = r^{\text{int}}_{x} + \frac{1}{N} \sum_{i=1}^{N} p_i \ast k_x$. The context representations are updated with a separate kernel $r^{\text{out}}_{C_i} = r^{\text{int}}_{C_i} + p_i \ast k_C$. We then re-size the outputs of a Pairwise-Conv-Avg Block before feeding them as input the next block. We experimented with attention-based and weighted average approaches but found that they did not lead to an increased generalization to unseen tasks.

\subsection{Task augmentations} \label{sec:task_augmentations}
To further diversify the training dataset, we employ task augmentations~\cite{universeg2022}, a group of transformations applied at random to the input, output, and context images. The objective is to increase the diversity of tasks to discourage the model from merely memorizing the tasks in the training data. A list of all task augmentations is summarized in \cref{tab:task_augmentations}, with more detailed descriptions and visual examples in Supplement~\ref{supplement:augmentations}.

\begin{table}[b]
\caption{Task Augmentations}
\label{tab:task_augmentations}
\centering
\begin{tabularx}{\linewidth}{Xl}
\toprule
Task Augmentations  &                        \\ \midrule
IntensityMapping    & SyntheticModality                 \\
SobelFilter   & MaskInvert           \\
MaskContour         & MaskDilation                \\
PermuteChannels         & DuplicateChannels                 \\\bottomrule
\end{tabularx}
\vspace{-1em}
\end{table}

\subsection{Inference}
During inference, we supply an input image $x_i$ and a context set $C_i$ from the desired task.
Given these inputs, a simple feed-forward pass through the model provides the prediction $\hat{y} = g(x, C)$. To further increase accuracy at test-time, we use context-set bootstrapping~\cite{universeg2022}. We also increase the context set by sampling with replacement from the context set, and add small affine augmentations.

\section{Experiments}
We first compare \name with task-specific networks, which require substantial expertise and compute. We then analyze the effect of the size of the context set, and the multi-task generalization to unseen segmentation protocols and image modalities. For this first method of large-scale multi-task generalization in neuroimaging, we conduct the experiments on 2D image slices.

\subsection{Data}
To create a diverse dataset encompassing a multitude of different modalities, acquisition protocols, devices, and tasks, we pool neuroimages from the public datasets OASIS3~\cite{Marcus2007,hoopes2022learning}, BRATS~\cite{Brats2015,Brats2017,Brats2018}, Atlas R2.0~\cite{Liew2022}, Hammers Atlas~\cite{Hammers2003}, IXI~\cite{IXI}, ISLES2022~\cite{Petzsche2022}, and the White Matter Hyperintensities Challenge~\cite{Kuijf2019}. We segment all subjects with Synthseg~\cite{Billot2021, fischl2012freesurfer}. Based on the segmentation, we affinely align the images to the MNI 152 template space~\cite{Fonov2009,Fonov2011}, and resample to 1mm isometric resolution at a size of $192 \times 224 \times 192$mm. We perform manual quality control of the segmentation and registration by ensuring no segmented areas fall outside of the cropped volume and discard subjects failing this check (4 subjects). We extract a coronal slice of $192 \times 192$mm, bisecting the frontal Brain stem, Hippocampus, Thalamus, and Lateral ventricles. We rescale image intensities to the $[0, 1]$ interval using dataset-specific percentiles. For full head images, we create a brain mask with Synthstrip~\cite{Hoopes2022, fischl2012freesurfer}. The final dataset contains 2,282 subjects with 15,911 images and segmentation masks across 8 modalities. Subjects of the seven original datasets are split into $80\%$ for training and validation, $20\%$ test, with a minimum of 15 test subjects per dataset.

\begin{figure*}
    \centering
    \includegraphics[width=1.0\linewidth]{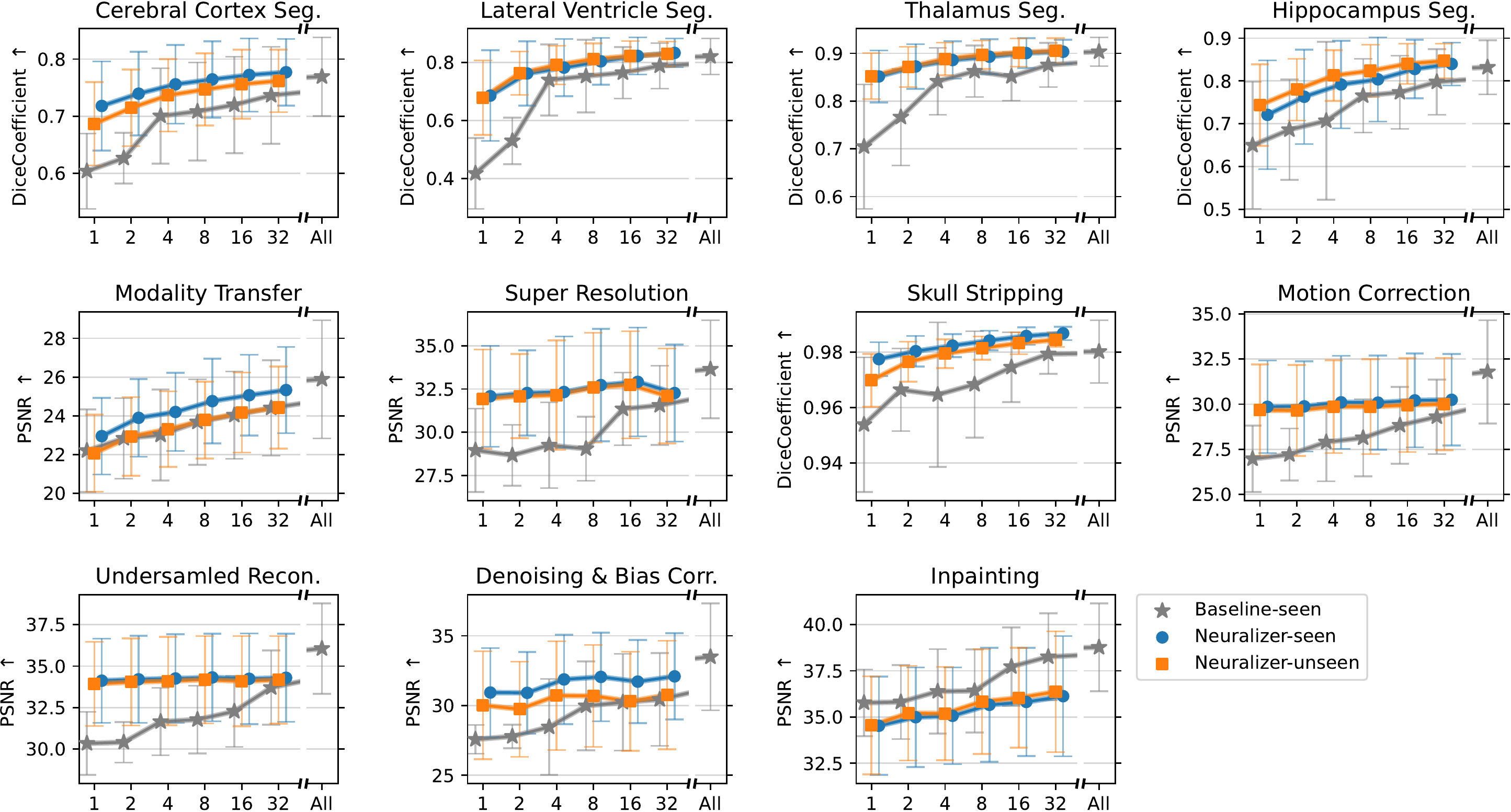}
    \caption{Performance of multi-task \name and the task-specific baselines on each task, averaged across all modalities in the test set. The tasks being evaluated were included in the training of \name-seen (orange), held out in \name-unseen (blue), and specifically trained on by each task-specific baseline (gray). The x-axis is the size of the train/context set, and the y-axis is the Dice/PSNR score. Some points on the x-axis are omitted for better visibility. `All' refers to all available train data for the task, ranging from 249 to 2,282 subjects depending on the task. The bars denote standard deviation across modalities. We extract results for T1 scans in Supplement~\ref{supplement:score_per_task_t1}.}
    \label{fig:performance_per_task}
\end{figure*}

\subsection{Models}

\paragraph{\name-seen.} This \name model includes all tasks available during training. We use this model to evaluate the performance on unseen scans from tasks and modalities that have been included in the training. The model uses the 4-stage architecture shown in \cref{fig:model} with 64 channels per layer. During training, the context size $|C_i|$ is sampled from $\mathcal{U}_{\{1, 32\}}$ at each iteration.

\paragraph{\name-unseen.}
To evaluate \name performance on tasks and modalities it has not been trained on, we train a family of \name models where a single task or modality is excluded from the training set. The model architecture of \name-unseen is identical to \name-seen.

\paragraph{Baseline-seen.}
As no established baseline for multi-task and multi-modality models in neuroimaging can tackle the number of tasks we aim for, we compare \name to an ensemble of task-specific U-Nets~\cite{Ronneberger2015,Milletari2016}. However, training one model for each task and modality requires overwhelming computational resources. To reduce the computational requirement, we follow previous modality-agnostic models~\cite{Billot2021, Hoopes2022} and train each model on multiple input modalities. This lowers the number of models to be trained to one per task, segmentation class, and modality-transfer output modality. To compare Baseline-seen with \name-unseen given an equal amount of data, we train baselines with training set sizes of $\{1,2,4,8,16,32,\text{all}\}$ and employ standard data augmentation.

We use a 4-stage U-Net architecture with one residual block per layer. The channel width is tuned experimentally for each training dataset size. We select 256 channels when all data is available for training, and 64 channels otherwise. Using larger U-Nets resulted in overfitting and lower performance. Supplement~\ref{supplement:inference_cost} summarizes model parameter counts and inference costs.

\begin{figure}
    \centering
    \includegraphics[width=1\linewidth]{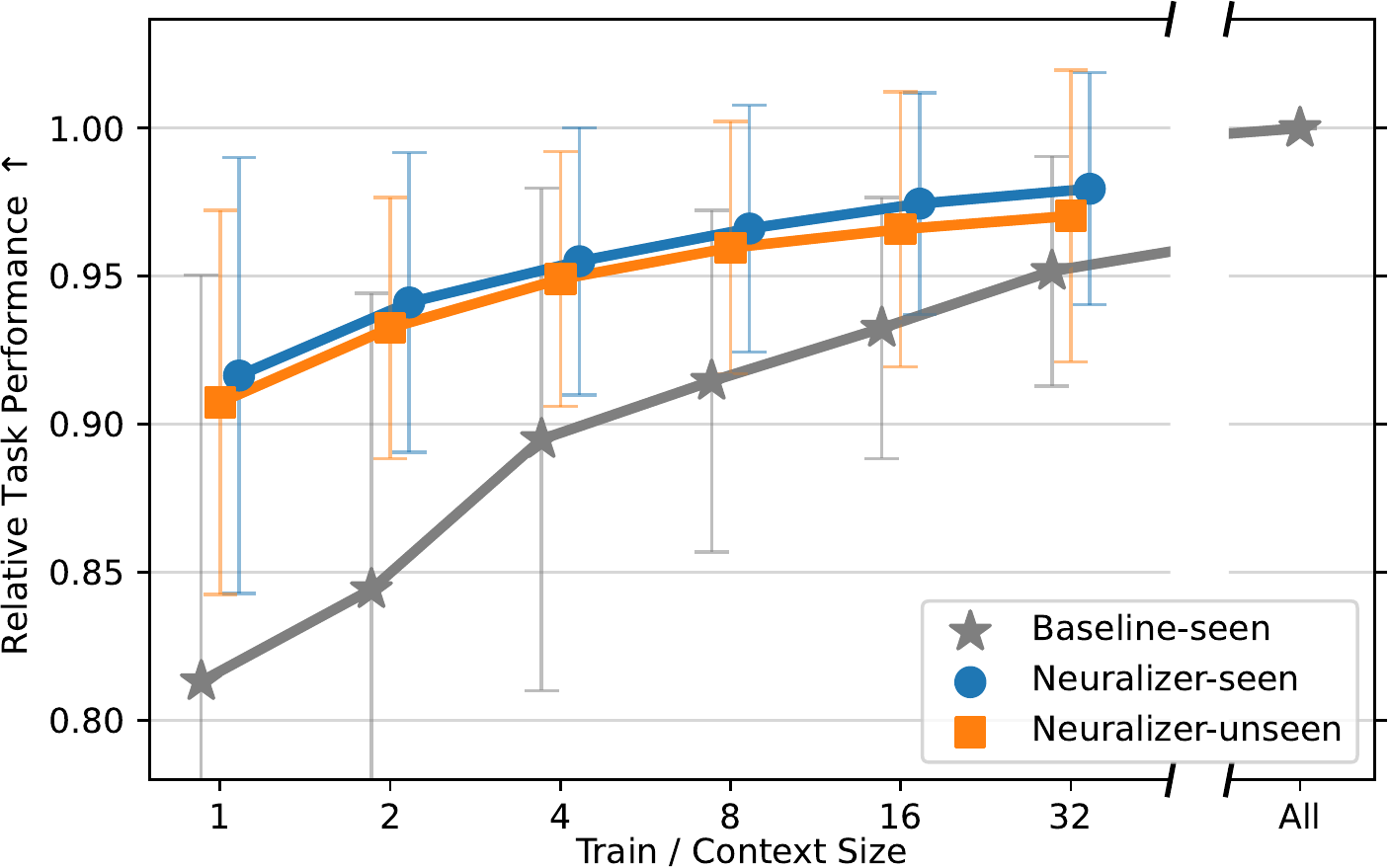}
    \caption{Results averaged across tasks, expressed as relative performance compared to the baseline trained on all data. The tasks being evaluated were included in the training of \name-seen (orange), held out in \name-unseen (blue), and specifically trained on by each task-specific baseline (gray). The x-axis is the size of the train/context set, and the y-axis is the relative score. Some points on the x-axis are omitted for better visibility. `All' refers to all available data for the task, ranging from 249 to 2,282 subjects depending on the task. Bars: standard deviation across tasks/modalities.}
    \vspace{-0.5em}
    \label{fig:rel_performance}
\end{figure}

\subsection{Training}
We use supervised training, task-specific loss functions, and weigh the MSE loss by selecting $\sigma^2 = 0.05$, resulting in the loss terms being of similar magnitude. All models are trained with a batch size of 8, a learning rate of $10^{-4}$, and the ADAM optimizer~\cite{Kingma2015}. To speed up training, we under-sample tasks that the model learns quickly, with sampling weights given in Supplement~\ref{supplement:task-weights}.

In addition to the task augmentations, we use data augmentations via random affine movements, random elastic deformations, and random flips along the sagittal plane. For Baseline-seen, we reuse the augmentations but remove those that introduce uncertainty in the desired output. The training time for \name is 7 days on a single A100 GPU. The training time of the baseline models is capped at 5 days. All models use early stopping, ending the training after 25 epochs exhibiting no decrease in validation loss. The model with the lowest validation loss is used for further evaluation on the test set. 

\subsection{Evaluation}
We evaluate the Dice coefficient for the segmentation and skull stripping tasks, and the Peak Signal-to-Noise Ratio (PSNR) for the image-to-image tasks on the test set. As the low-data regime is of particular interest, we measure performance as a function of context set size for the \name models and use training set size as an analog for the U-Net models. We evaluate context sets of up to 32 subjects. Larger context sets are possible but come at a linear cost in memory.


\subsection{Experiment 1: Baseline Comparison}
To assess if the proposed multi-task approach is competitive with task-specific models, we evaluate the performance and runtime of \name-seen, \name-unseen, and Baseline-seen on the test-set of each task. For \name-unseen, we withhold image modalities using a leave-one-out strategy during training and evaluate on the unseen modalities at test time.

\vspace{-0.2cm}  \paragraph{Results.}
We display the results by task, averaged across modalities in \cref{fig:performance_per_task}. We also provide an evaluation of using just the T1 modality in Supplement~\ref{supplement:score_per_task_t1}, since many task-specific networks in neuroimage analysis literature focus on T1 images. We further aggregate performance across all tasks in \cref{fig:rel_performance}, and provide tabular results in Supplement~\ref{supplement:experiment_results}.
Both \name models outperform most task-specific baselines trained on up to 32 samples. When training the baselines on all available data, the baselines outperform \name-seen by 2 percentage points in relative performance, and \name-unseen by 3 percentage points.
The loss in performance when generalizing to an unseen modality (between \name-seen and \name-unseen) is less than 2 percentage points for all context set sizes. 

Training the baseline model to convergence on 32 samples took on average $28.2 \pm 16.6$ hours per task, using one A100 GPU. Since \name only requires inference for a new task, it is orders of magnitude faster, requiring less than $0.1$ seconds on a GPU and less than $3$ seconds on a CPU.

We provide qualitative samples of the predictions from \name-seen model in Supplement~\ref{supplement:samples}, Figures~\ref{fig:appendix_samples_1}-\ref{fig:appendix_samples_3}. 

\subsection{Experiment 2: Context set size analysis}
We assess the few-shot setting that is prevalent in neuroimage analysis, where few annotated images are often available for a new task. We evaluate performance as a function of the number of labeled samples. For \name, we evaluate the model with context-set sizes of $\{1,2,4,8,16,32\}$ unique subjects from the test set. For the baseline, we trained models with reduced training set sizes of the same amount of subjects. To reduce the effect of random  training subject selection, we train three separate baselines with $n=1$, two baselines with $n=2$, and average results of models with the same $n$. 

\vspace{-0.2cm}  \paragraph{Results.}
\cref{tab:score_per_task} and Figs.~\ref{fig:performance_per_task},~\ref{fig:rel_performance} illustrate the results. For all models, prediction accuracy increases with the availability of labeled data, with diminishing returns. For both \name models, a context set size of one achieves more than 90\% of the performance attainable with all data. For most tasks, the baseline performs overall worse than both \name models when $\leq 32$ labeled samples are available but achieves the best overall performance on larger datasets.

\begin{table*}[]
    \centering
    \setlength{\tabcolsep}{3pt}
\begin{tabularx}{\linewidth}{Xcccccccccccccccc}
\toprule
     Model & Task Seen &   \multicolumn{14}{c}{Segmentation Class (Hammers Atlas)} & Mean \textit{(std)} \\ \cmidrule(lr){3-16}
     & &   \multicolumn{1}{c}{Hip} &  \multicolumn{1}{c}{PAG} &  \multicolumn{1}{c}{STG} &  \multicolumn{1}{c}{MIG} &  \multicolumn{1}{c}{FuG} &  \multicolumn{1}{c}{Stm} &  \multicolumn{1}{c}{Ins} &  \multicolumn{1}{c}{PCG} &  \multicolumn{1}{c}{Tha} &  \multicolumn{1}{c}{CC} &  \multicolumn{1}{c}{3V} &  \multicolumn{1}{c}{PrG} &  \multicolumn{1}{c}{PoG} &  \multicolumn{1}{c}{ALG} & \\
\midrule
       Baseline-seen &          \cmark & .88 & .86 & .93 & .92 & .79 & .87 & .82 & .87 & .90 & .80 & .68 & .83 & .77 & .82 &  .84 \textit{(.07)}\\
  \name-seen &          \cmark & .88 & .86 & .92 & .92 & .76 & .88 & .83 & .85 & .90 & .82 & .73 & .86 & .77 & .81 &  .84 \textit{(.07)}\\
\name-unseen &          \xmark & .88 & .87 & .93 & .91 & .78 & .87 & .82 & .85 & .90 & .81 & .72 & .85 & .78 & .81 &  .84 \textit{(.06)}\\
\bottomrule
\end{tabularx}
    \caption{Segmentation of the Hammers Atlas dataset. For \name-unseen, this dataset and segmentation protocol is withheld from training. Evaluation of major labels located in the center and right of the coronal slice. See Supplement~\ref{supplement:hammers_atlas_class_names} for class abbreviations.}
    \label{tab:hammers_atlas}
\end{table*}

\subsection{Experiment 3: Generalization to a new segmentation protocol}
The Hammers Atlas dataset~\cite{Hammers2003,Gousias2008,Faillenot2017} provides an alternative anatomical segmentation protocol to the widely-used Freesurfer segmentation available for most subjects in the dataset. The shape, size, and amount of annotated regions in the protocols differ drastically. A different image acquisition site also leads to differences in visual characteristics. We use the Hammers Atlas dataset to evaluate \name-unseen by entirely withholding the dataset and its annotations from training.  We evaluate the Dice coefficient of the 14 major anatomical segmentation classes present in the center and right half of the coronal slice.

\vspace{-0.2cm} \paragraph{Results.}
\cref{tab:hammers_atlas} illustrates the results. \name-unseen performs similarly to \name-seen and the baseline, while not requiring lengthy re-training or fine-tuning on the Hammers Atlas dataset, and not having seen the segmentation protocol. All three models achieve a mean Dice coefficient of $0.84$. The largest performance difference is in the third ventricle class, where both \name models outperform the baseline by at least $0.04$ Dice. The Freesurfer segmentation protocol included in the training set of the \name models also contains a third ventricle class.

\section{Discussion}
Our experiments using modality and segmentation class hold-outs show that \name can generalize well to unseen neuroimaging tasks. Across all context set sizes, the generalization loss between seen and unseen modalities and segmentation classes is less than $2$ percentage points across Experiments 1 and 2. On the smaller held-out Hammers-Atlas segmentation dataset, we find that \name can generalize to unseen tasks with similar performance. These results show promise that a single \name model can perform multiple neuroimaging tasks, including generalization to new inference tasks not seen during training.

In settings with 32 or fewer labeled example images, \name-unseen outperforms task-specific baselines despite never having seen the task or modality at train time, and taking nearly no effort or compute compared to the baselines which require substantial expertise, manual labor, and compute resources. The performance difference is largest when only one labeled subject is available, but still present at 32 subjects~(\cref{fig:rel_performance}). \name provides a performance advantage on smaller datasets likely by exploiting neuroimaging similarities across the many other neuroimaging tasks and datasets available in training. 

When training the baselines on all available data, they can outperform \name-seen and \name-unseen by at most 3 percentage points. The inflection point of identical performance between \name and the baselines is not covered by the range of context set sizes chosen for training and evaluation due to prohibitive computational costs and is an interesting direction of study.
When large annotated datasets are available, the baselines performed best on most tasks. However, training task-specific models comes at a significant cost. As a first step in the proposed problem formulation, \name offers an alternative with near equal performance, while only requiering seconds to infer any task from the context set.

\subsection*{Limitations}
We made simplifying assumptions in this first paper demonstrating the potential of multi-task generalization in neuroimaging.
The experiments are conducted on 2D data slices. In large part, we did this since running the hundreds of baselines in 3D would be infeasible on our compute cluster. Entire volumetric data also impose a challenging memory requirement on \name models. To tackle 3D data in the future, we plan to process multiple slices at a time.

We affinely aligned the neuroimages of the context set to the target image. Early in \name development, we tried training on non-aligned inputs but found that it deteriorated performance. The need for affine alignment provides an obstacle to adoption. While existing affine-alignment tools are fast and can be employed, we also believe that this requirement can be removed with further development.

We originally experimented with tumor and lesion segmentation tasks but found this to be a more challenging scenario. Lesions are spatially heterogeneous, making learning from the context set much harder. We excluded tumor and lesion segmentation masks from the experiments, but plan to study this setting in the future.

While we demonstrate the proposed ideas on a broad range of tasks and modalities, neuroimage analysis can involve more domains, tasks, and populations, like image registration, surface-based tasks, CT image domains, and pediatric data. We plan to extend \name to tackle these in the future.

\section{Conclusion}
\name performs accurate rapid single-pass, multi-task generalization, and 
 even outperforms task-specific baselines in limited data scenarios. Even when a large amount of annotated data is available, \name often matches baseline performance despite not training on the data. \name provides clinical researchers and scientists with a single model to solve a wide range of neuroimaging tasks on images of many modalities, and can be easily adapted to new tasks without the prohibitive requirement of retraining or fine-tuning a task-specific model. We believe this will facilitate neuroscience analyses not currently possible.

\section*{Acknowledgments}
This work was performed during Steffen Czolbes's visit to the Athinoula A. Martinos Center for Biomedical Imaging, MGH. The collaboration was supported by the EliteForsk Travel Grant of the Danish Ministry of Higher Education and Science. The work was further funded by the Novo Nordisk Foundation (grants no.~NNF20OC0062606 and NNF17OC0028360), the Lundbeck Foundation (grant no.~R218-2016-883), and NIH grants 1R01AG064027 and 1R01AG070988.

{\small
\bibliographystyle{ieee_fullname}
\bibliography{neuralizer}
}

\clearpage
\onecolumn
\appendix
\begin{appendices}

\section{Samples} \label{supplement:samples}
We provide examples of model inputs -- target image and context set -- and \name-seen predicted outputs. The inputs are sampled at random from the test dataset. The context set length is sampled from the discrete random uniform distribution $\mathcal{U}_{\{1, 32\}}$. To reduce visual clutter, we display up to eight context image pairs and omit the rest in the visualization. We also only show one channel, excluding additional inputs like multiple modalities, or the binary mask for in-painting tasks. We provide a collection of images from the first 50 samples from the test dataset. We only excluded examples to avoid duplication of tasks.

\begin{figure*}[b]
    \centering
    \includegraphics[width=0.5\linewidth]{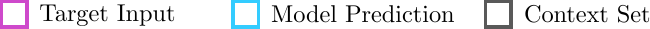}
    \includegraphics[width=1\linewidth]{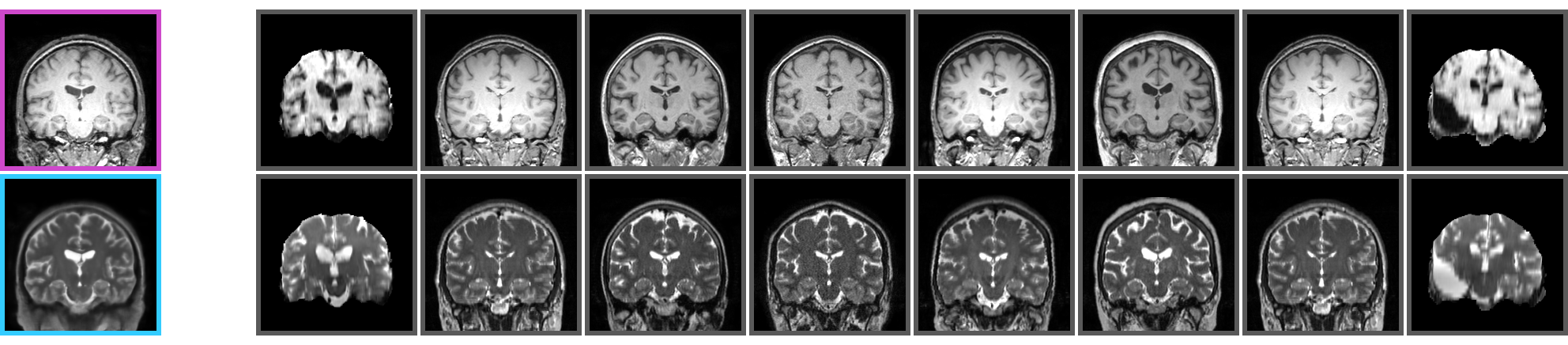}
    \includegraphics[width=1\linewidth]{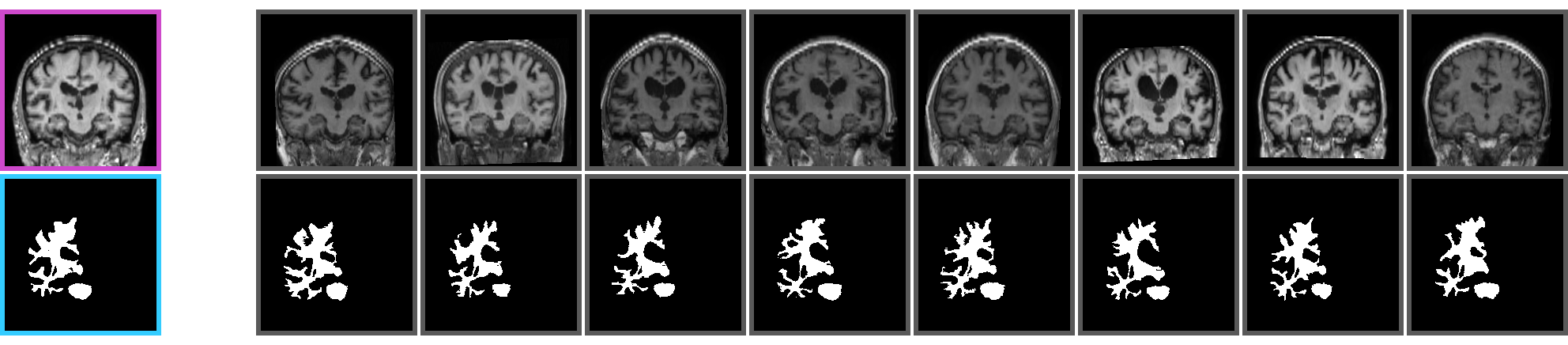}
    \includegraphics[width=1\linewidth]{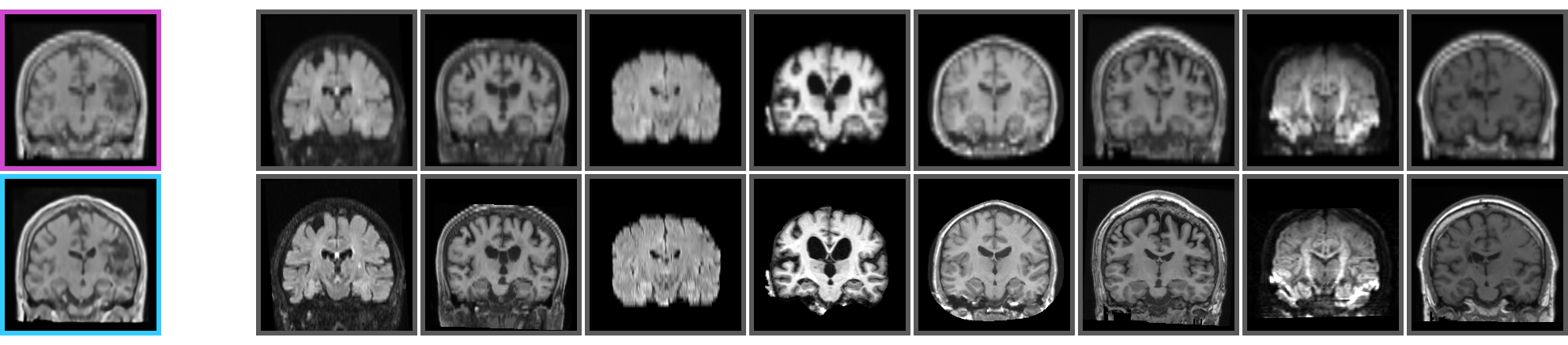}
    \caption{Sample \name-seen predictions. Left: Target input (magenta frame) and model prediction (blue frame). Right: context set supplied to inform the task (grey frame). We provide more samples on the next pages.}
    \label{fig:appendix_samples_1}
    \vspace{3cm}
\end{figure*}

\begin{figure*}[h!]
    \centering
    \includegraphics[width=0.5\linewidth]{img/samples_legend.pdf}
    \includegraphics[width=1\linewidth]{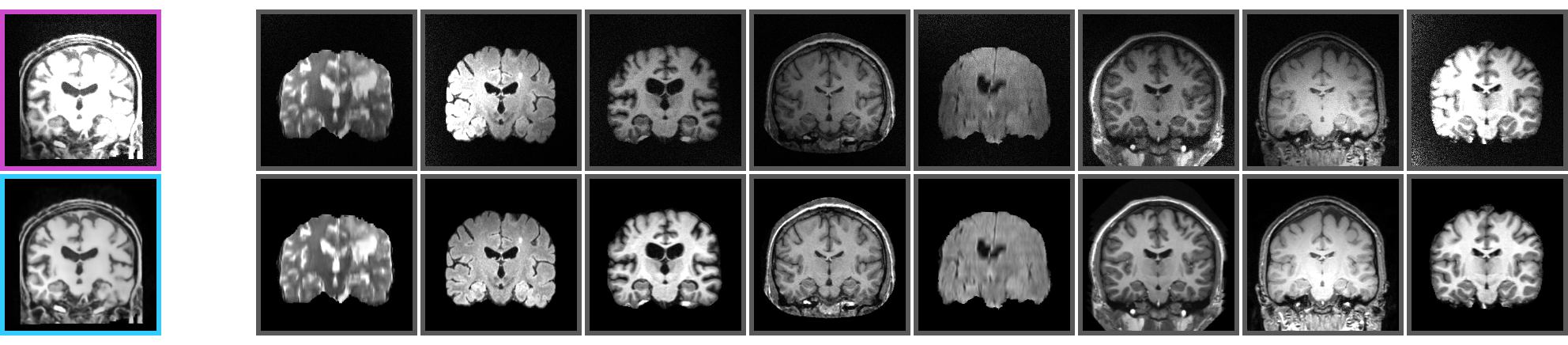}
    \includegraphics[width=1\linewidth]{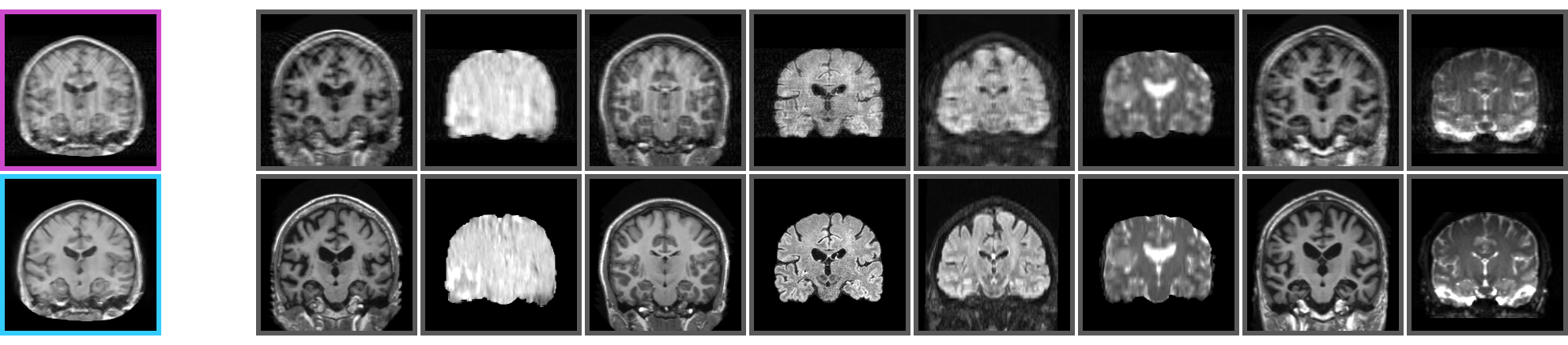}
    \includegraphics[width=1\linewidth]{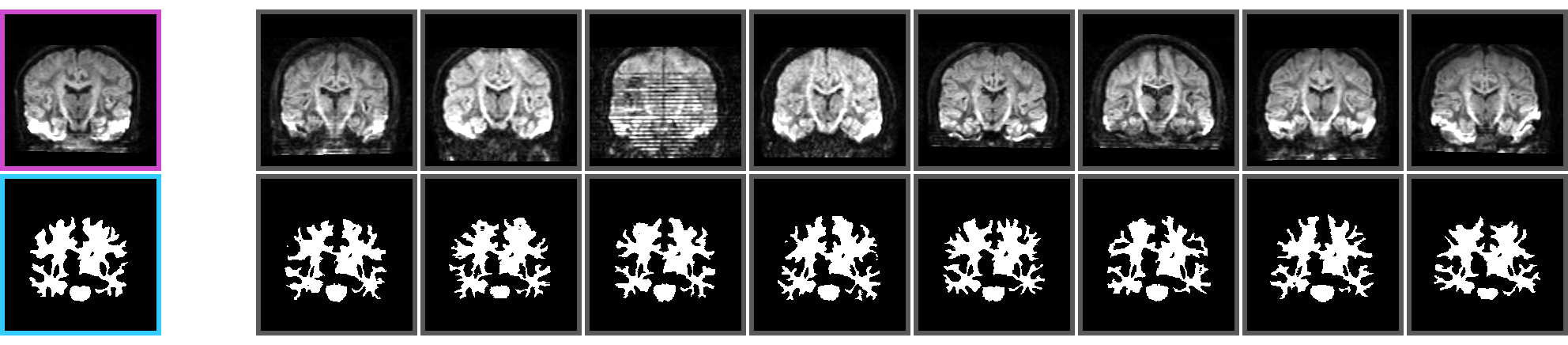}
    \includegraphics[width=1\linewidth]{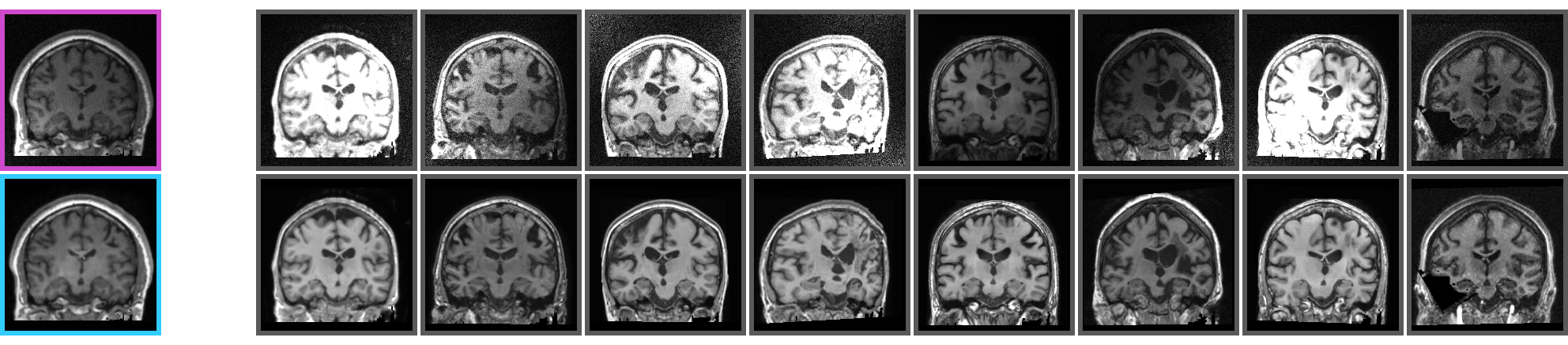}
    \includegraphics[width=1\linewidth]{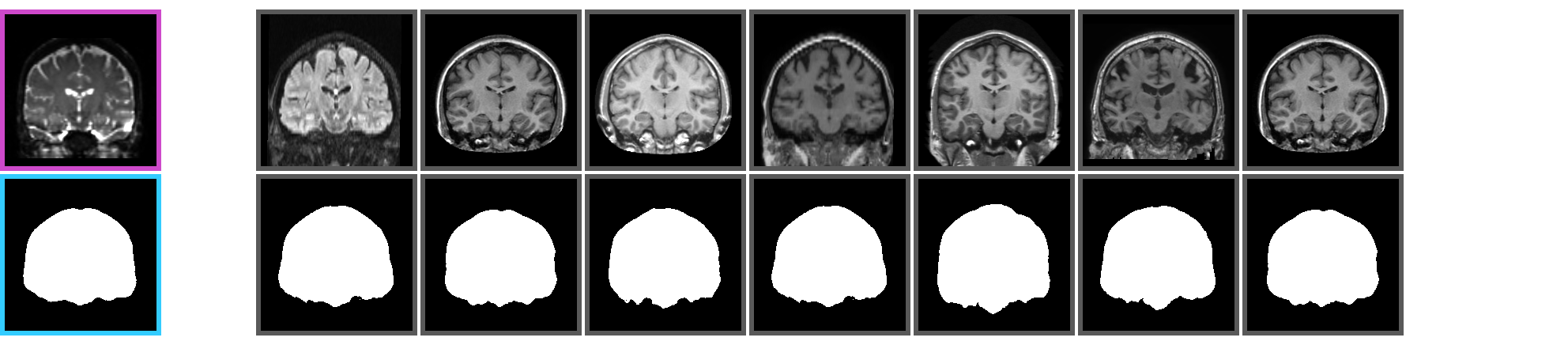}
    \caption{Sample \name-seen predictions (continued). Left: Target input (magenta frame) and model prediction (blue frame). Right: context set supplied to inform the task (grey frame).}
    \label{fig:appendix_samples_2}
\end{figure*}

\begin{figure*}[h!]
    \centering
    \includegraphics[width=0.5\linewidth]{img/samples_legend.pdf}
    \includegraphics[width=1\linewidth]{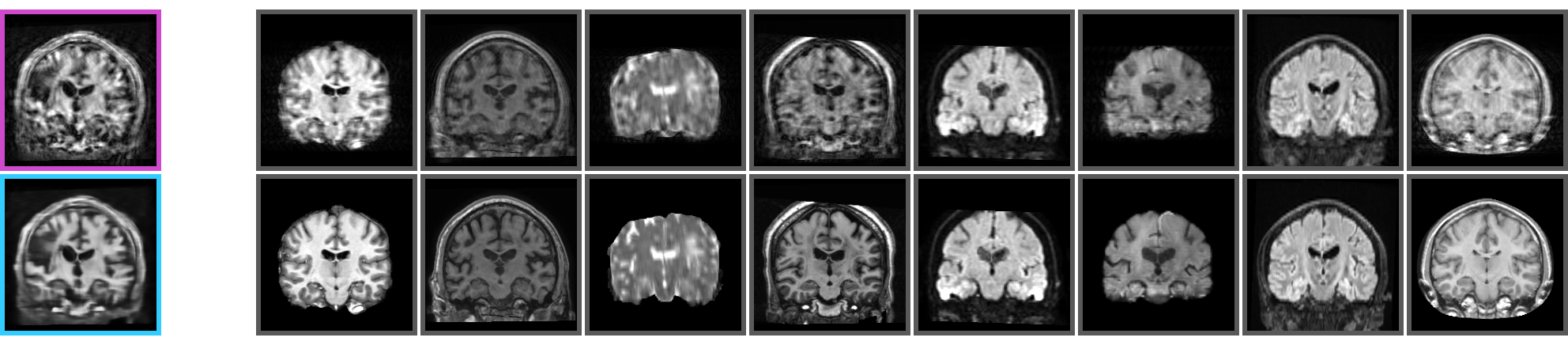}
    \includegraphics[width=1\linewidth]{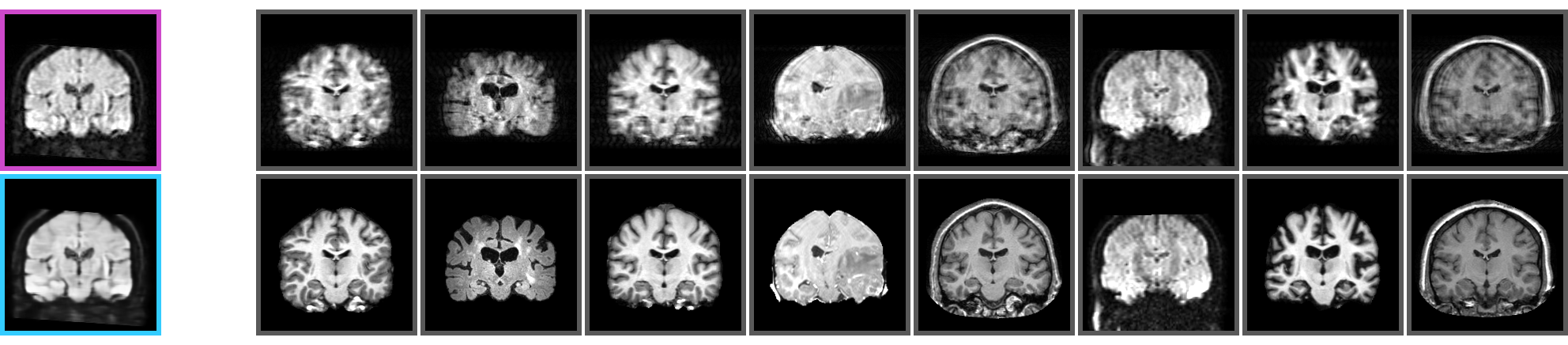}
    \includegraphics[width=1\linewidth]{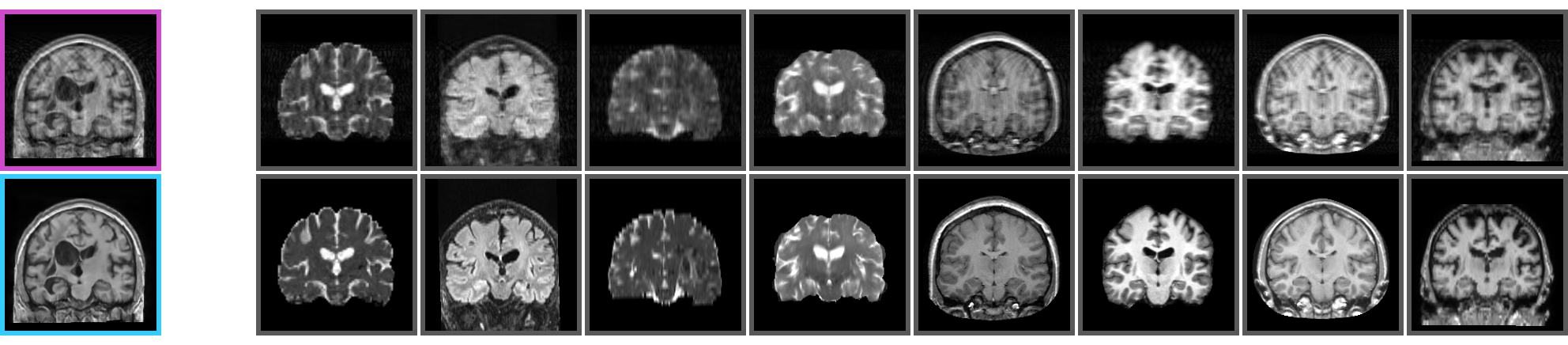}
    \includegraphics[width=1\linewidth]{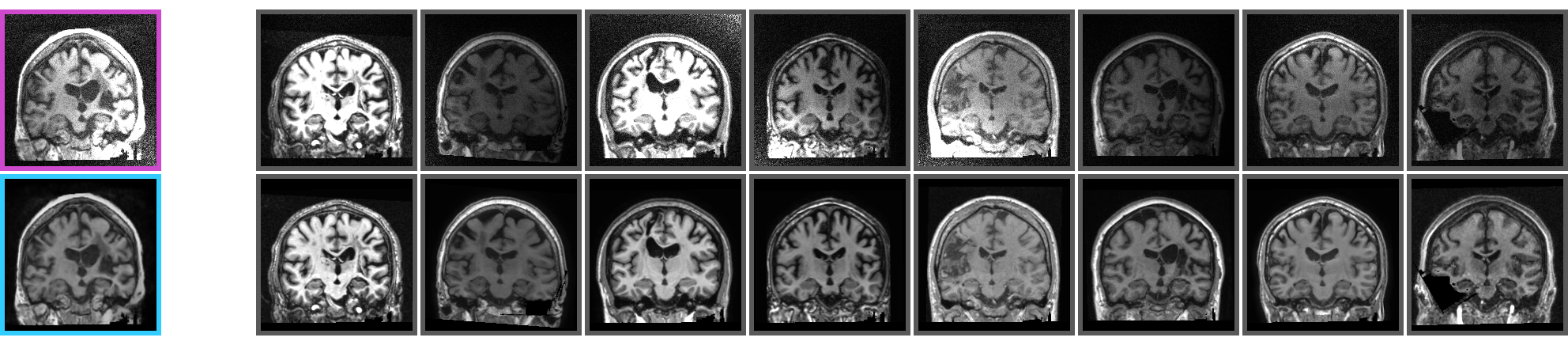}
    \includegraphics[width=1\linewidth]{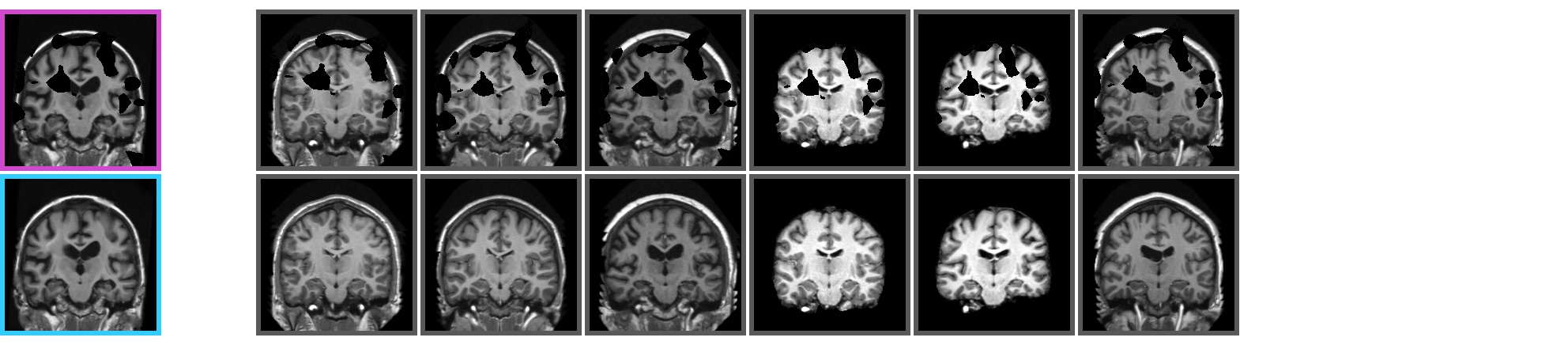}
    \caption{Sample \name-seen predictions (continued). Left: Target input (magenta frame) and model prediction (blue frame). Right: context set supplied to inform the task (grey frame).}
    \label{fig:appendix_samples_3}
\end{figure*}

\clearpage
\section{Train samples} \label{supplement:experiment_results}
We provide samples from the train set, including data and task augmentations, and show all three input channels. Further examples of the visual diversity possible with task augmentations are shown in \cref{fig:augmentaions}.

\begin{figure*}[b]
    \centering
    \includegraphics[width=0.5\linewidth]{img/samples_legend.pdf}
    \includegraphics[width=1\linewidth]{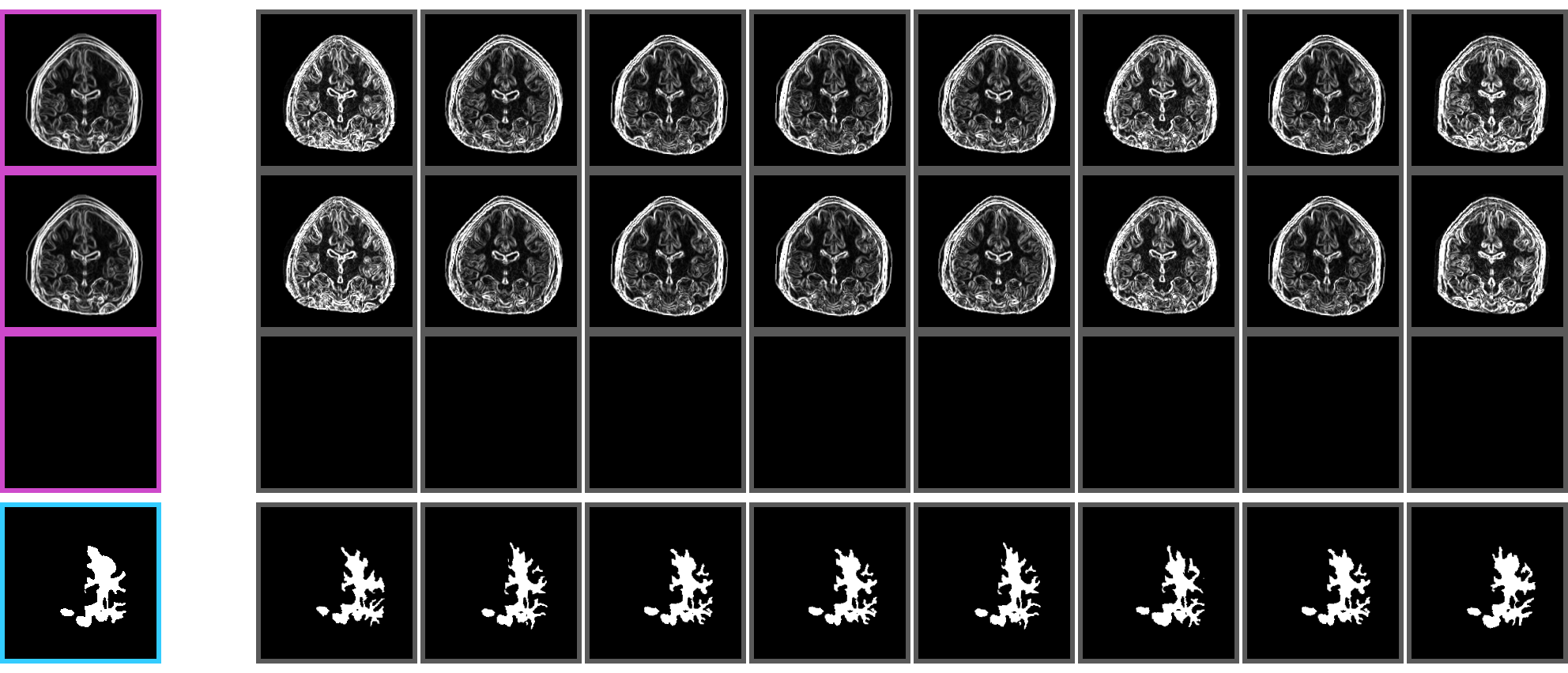}
    \includegraphics[width=1\linewidth]{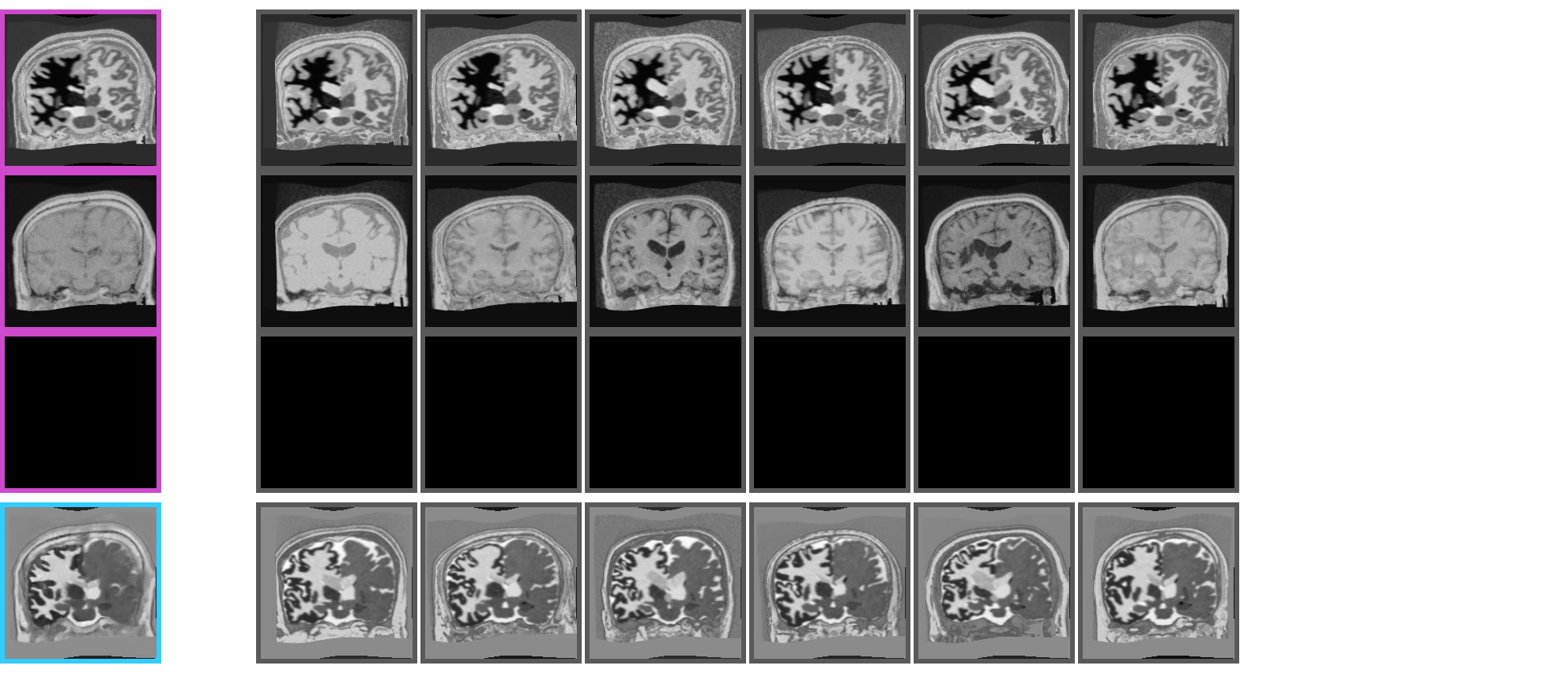}
    \caption{Sample \name-seen predictions from the train set, with data and task augmentations. All three channels of the input are shown. Left: Target input (magenta frame) and model prediction (blue frame). Right: context set supplied to inform the task (grey frame).}
    \label{fig:appendix_train_samples}
    \vspace{3cm}
\end{figure*}

\clearpage
\section{Task augmentations} \label{supplement:augmentations}
Task augmentations are randomized data augmentations applied to both input and target images or segmentation maps. These change not only the appearance of the input image, but also the target and members of the context set, essentially altering the task itself.  We apply task augmentations not to create plausible neuroimaging tasks, but instead to expand the set of tasks the model is exposed to during training. This prevents memorization of the training tasks, and aids generalization to unseen tasks during inference. We first describe the task augmentations in \ref{supplement:augmentations1}, then discuss their composition in \ref{supplement:augmentations2}, and finally provide examples in \cref{fig:augmentaions}. Hyper-parameters for all augmentations are selected by visual inspection.

\subsection{Task augmentations} \label{supplement:augmentations1}
We provide a description of each task augmentation. In addition to the task augmentations, we use data augmentations via random affine movements, random elastic deformations, and random flips along the sagittal plane.

\paragraph{SobelFilter.}
A Sobel filter is applied to an intensity image.

\paragraph{IntensityMapping.}
The intensity of an image is remapped~\cite{Hoffmann2020a} To perform this operation, the image intensity values are split into histogram bins, and each bin is assigned a new intensity difference value. To obtain new intensity values, we compute a distance from the original intensity value to the two neighboring bin centers, using linear interpolation.

\paragraph{SyntheticModality.}
An intensity image is replaced with a synthetic one generated from an anatomical segmentation map of the subject, using previous work~\cite{Hoffmann2020a}. Each anatomical segmentation class is randomly assigned an intensity mean and standard deviation and the new synthetic modality image of the brain is generated according to these distributions. As our anatomical segmentations do not cover the skull, we take an extra step to ensure skulls are present in the synthetic data: If the original intensity image had a skull, the generated brain is overlaid onto the original image, thus keeping the skull.

\paragraph{MaskContour.}
We extract a contour of the binary mask in a segmentation task, which then represents the new target segmentation mask. Contoured Masks are always dilated to a width of 3 voxels.

\paragraph{MaskDilation.}
The binary segmentation mask is dilated by 1 voxel.

\paragraph{MaskInvert.}
The binary segmentation mask is inverted.

\paragraph{PermuteChannels.}
The input images are represented by three channels. On each input during training, we permute the input channels. This encourages the network to ignore the specific channel order.

\paragraph{DuplicateChannels.}
We overwrite empty input channels with the duplication of a non-zero channel. The augmentation is applied to each empty channel with a probability $p$.

\subsection{Composition and likelihood of task augmentations} \label{supplement:augmentations2}
We compose task and data augmentations during training. Some task augmentations can be combined (e.g. MaskDilation and MaskInvert), while others are exclusive to each other (e.g. SobelFilter and SyntheticModality). To model these dependencies, we define the default composition tree used for most tasks in \cref{supplement:augmentaion_composition}. The augmentation groups ``Mask Augmentations'', ``Intensity Augmentations'', ``Channel Augmentations'', and ``Spatial Augmentations'' are applied in this order. Augmentations in child nodes of ``Compose'' are applied left to right, while ``OneOf'' selects a single child augmentation to apply. A node is applied with probability $p$ stated on the node.

Some tasks use modified versions of this composition tree. As a safety feature, we do not use RandomFlip for segmentation-related tasks, as this can lead to information leakage when evaluating on non-symmetric class-holdouts (in our experiments presented here we always hold out the same anatomical class on both sides of the brain, but this has not always been the case during development). To simplify other tasks, we omit MaskContour and MaskDilate from the inpainting task, and SobelFilter and SyntheticModality form the modality transfer task.

\begin{figure*}[h]
  \centering
    \includegraphics[width=1\linewidth]{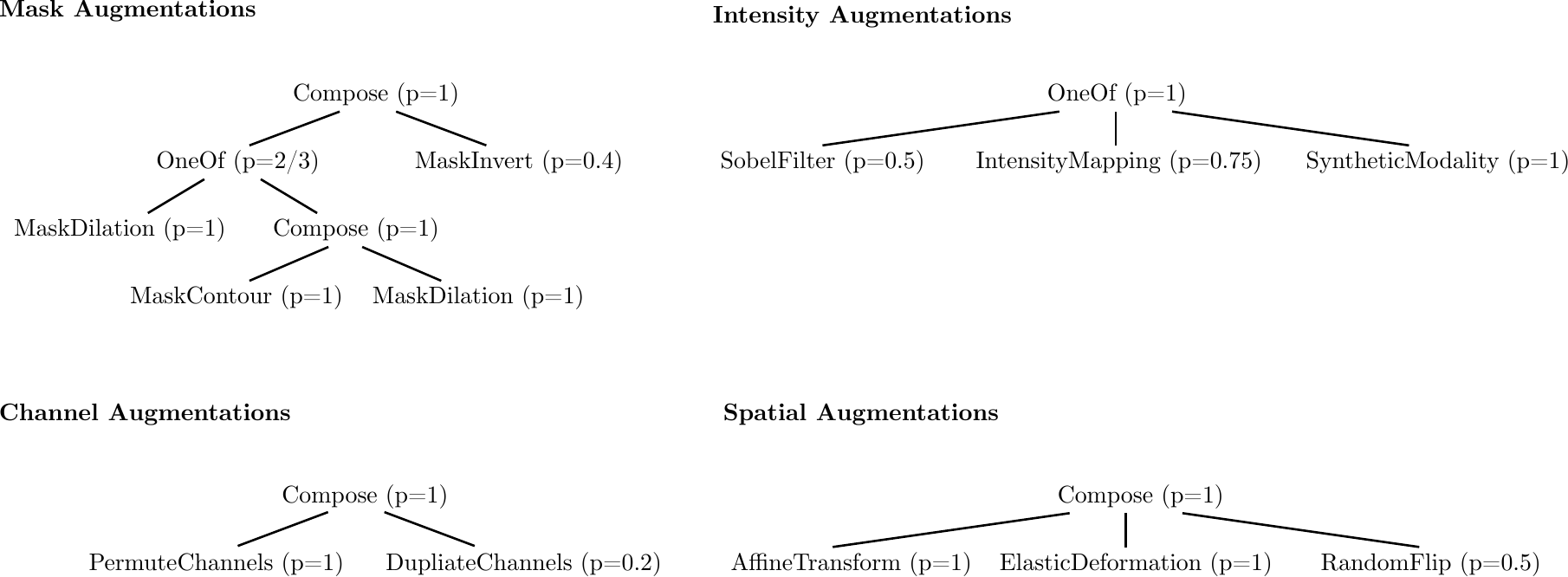}

   \caption{Default composition of augmentations used for most tasks during training. We use ``Compose'' and ``OneOf'' nodes to model these restrictions. Augmentations in child nodes of ``Compose'' are applied left to right, while ``OneOf'' selects a single child augmentation to apply. A node is applied with probability $p$.}
   \label{supplement:augmentaion_composition}
\end{figure*}

\subsection{Examples of task augmentations}
\begin{figure*}[hb]
  \centering
    \includegraphics[width=1\linewidth]{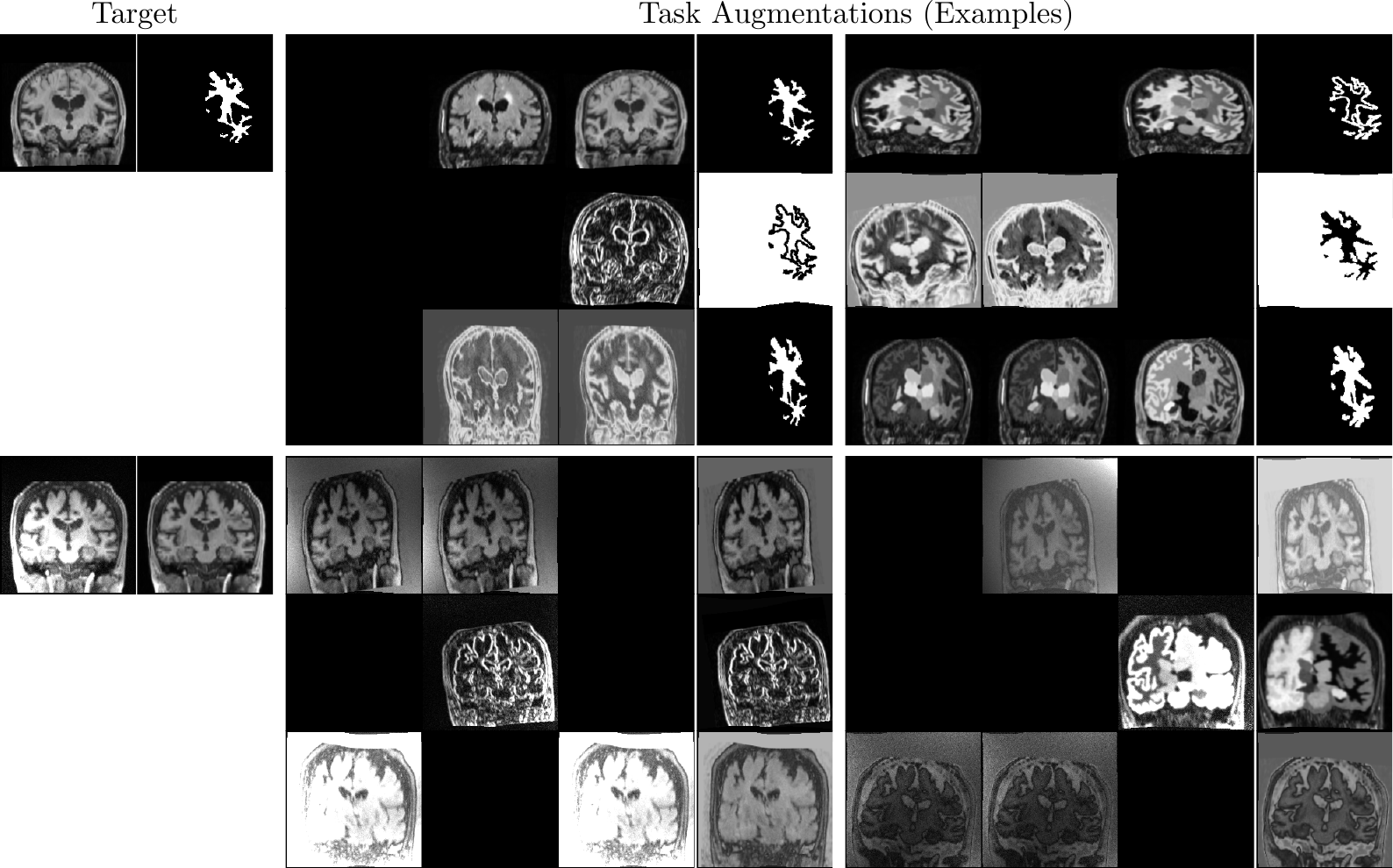}

   \caption{Examples of task augmentations, designed to increase the diversity of neuroimaging tasks seen by the model during training. We show non-augmented target input and output image of T1 modality on the left. We show examples of random data- and task-augmentations applied to the target during training on the right. The augmented target input is represented by up to three channels of real and synthetic modalities of the subject. The target output is augmented with synthetic image modalities and alterations to the segmentation mask. The same augmentations are applied to the context set.}
   \label{fig:augmentaions}
\end{figure*}

\cref{fig:augmentaions} provides visual examples of task augmentations applied to a segmentation and bias correction task.

\clearpage
\section{Evaluation on T1 modality} \label{supplement:score_per_task_t1}
We aggregated scores across all modalities in \cref{fig:performance_per_task}. To aid comparison to existing literature, which most often focuses on T1 images, we provide the same evaluation, performed on just the T1 modality here. Some tasks are easier on T1 data, thus improving scores. For small dataset sizes of 1 or 2 subjects, the baselines sometimes underperform on the T1 modality. This is often because images of the T1 modality may not always present in small training sets. For sizes of 4 subjects and larger, the T1 modality is always included in the training set.
\begin{figure}[h]
    \centering
    \includegraphics[width=1.0\linewidth]{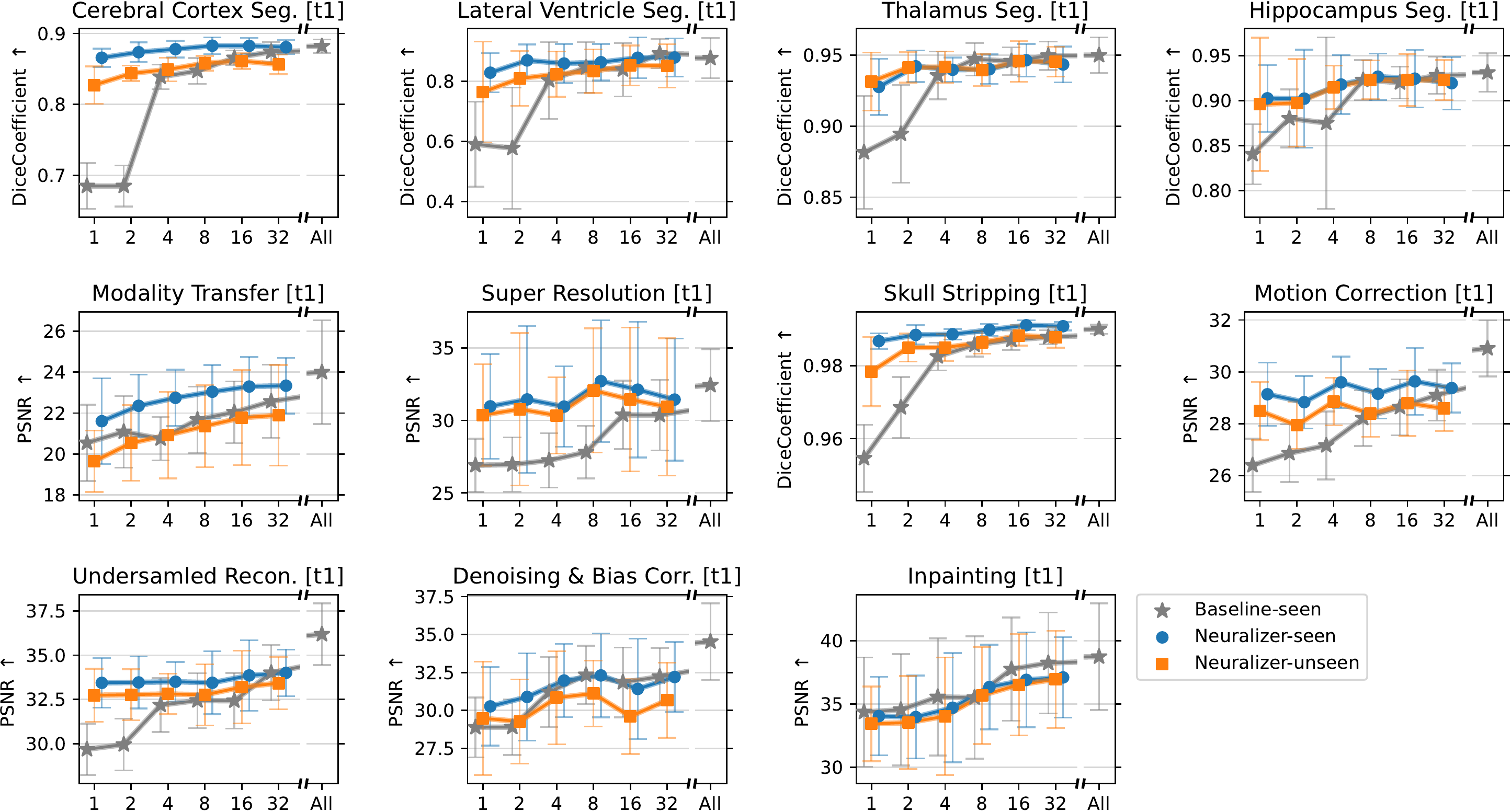}
    \caption{Performance of multi-task \name and the task-specific baselines on each task, T1 modality only. The tasks being evaluated were included in the training of \name-seen (orange), held out in \name-unseen (blue), and specifically trained on by each task-specific baseline (gray). The x-axis is the size of the train/context set, and the y-axis is the Dice/PSNR score. Some points on the x-axis are omitted for better visibility. `All' refers to all available train data for the task, ranging from 249 to 2,282 subjects depending on the task. The bars denote the standard deviation across subjects.}
\end{figure}

\clearpage
\section{Experiments 1 and 2 tabular results} \label{supplement:experiment_results}

\begin{table*}[h!]
    \centering
    \setlength{\tabcolsep}{4.5pt}    
\begin{tabularx}{\textwidth}{Xcrllllllll}
\toprule
\rot{Model} & \rot{Trained} & \rot{Subjects} & \rot{Segmentation} & \rot{Mod. Transfer} & \rot{Super Res.} & \rot{Skull Strip.} & \rot{Motion Recon.} & \rot{Undersamp. Recon.} & \rot{Noise Recon.} & \rot{Inpainting} \\
\midrule
     \multirow{7}{*}{\textvert{Baseline-seen}} &             \multirow{7}{*}{\cmark}  &            all &      $.83 \pm .08$ &          $25.9 \pm 3.0$ &   $33.6 \pm 2.8$ &         $.98 \pm .01$ &      $31.8 \pm 2.9$ &           $36.1 \pm 2.7$ &     $33.5 \pm 3.8$ &   $38.8 \pm 2.4$ \\
     &         &             32 &      $.80 \pm .09$ &          $24.4 \pm 2.5$ &   $31.6 \pm 2.3$ &         $.98 \pm .01$ &      $29.3 \pm 2.1$ &           $33.7 \pm 2.2$ &     $30.4 \pm 3.3$ &   $38.3 \pm 2.4$ \\
     &         &             16 &      $.78 \pm .09$ &          $24.0 \pm 2.3$ &   $31.3 \pm 2.1$ &         $.97 \pm .01$ &      $28.8 \pm 2.1$ &           $32.3 \pm 2.1$ &     $30.2 \pm 3.5$ &   $37.7 \pm 2.1$ \\
     &         &              8 &      $.77 \pm .11$ &          $23.7 \pm 2.2$ &   $29.0 \pm 1.9$ &         $.97 \pm .02$ &      $28.1 \pm 2.1$ &           $31.8 \pm 2.0$ &     $30.0 \pm 3.2$ &   $36.4 \pm 2.2$ \\
     &         &              4 &      $.75 \pm .14$ &          $23.0 \pm 2.3$ &   $29.2 \pm 2.5$ &         $.96 \pm .03$ &      $27.9 \pm 2.2$ &           $31.7 \pm 2.0$ &     $28.5 \pm 3.4$ &   $36.4 \pm 2.3$ \\
     &         &              2 &      $.65 \pm .12$ &          $22.8 \pm 2.1$ &   $28.7 \pm 1.8$ &         $.97 \pm .01$ &      $27.2 \pm 1.4$ &           $30.4 \pm 1.2$ &     $27.8 \pm 0.8$ &   $35.8 \pm 2.0$ \\
     &         &              1 &      $.59 \pm .16$ &          $22.2 \pm 2.1$ &   $29.0 \pm 2.4$ &         $.95 \pm .02$ &      $27.0 \pm 1.8$ &           $30.3 \pm 1.9$ &     $27.6 \pm 1.0$ &   $35.8 \pm 1.8$ \\ \addlinespace
  \multirow{6}{*}{\textvert{\name-seen}} &             \multirow{6}{*}{\cmark} &             32 &      $.84 \pm .07$ &          $25.3 \pm 2.2$ &   $32.3 \pm 2.8$ &         $.99 \pm .00$ &      $30.2 \pm 2.5$ &           $34.3 \pm 2.7$ &     $32.1 \pm 3.1$ &   $36.1 \pm 3.2$ \\
   &         &             16 &      $.83 \pm .07$ &          $25.1 \pm 2.1$ &   $32.9 \pm 3.1$ &         $.99 \pm .00$ &      $30.2 \pm 2.6$ &           $34.2 \pm 2.7$ &     $31.7 \pm 3.0$ &   $35.8 \pm 2.9$ \\
   &         &              8 &      $.82 \pm .09$ &          $24.8 \pm 2.2$ &   $32.7 \pm 3.3$ &         $.98 \pm .00$ &      $30.1 \pm 2.6$ &           $34.3 \pm 2.6$ &     $32.1 \pm 3.2$ &   $35.7 \pm 3.1$ \\
   &         &              4 &      $.80 \pm .09$ &          $24.2 \pm 2.0$ &   $32.3 \pm 3.2$ &         $.98 \pm .00$ &      $30.1 \pm 2.6$ &           $34.3 \pm 2.7$ &     $31.9 \pm 3.2$ &   $35.1 \pm 2.6$ \\
   &         &              2 &      $.78 \pm .10$ &          $23.9 \pm 2.0$ &   $32.3 \pm 2.5$ &         $.98 \pm .01$ &      $29.9 \pm 2.5$ &           $34.2 \pm 2.6$ &     $30.9 \pm 2.9$ &   $35.0 \pm 2.7$ \\
   &         &              1 &      $.74 \pm .13$ &          $23.0 \pm 2.0$ &   $32.1 \pm 2.9$ &         $.98 \pm .01$ &      $29.9 \pm 2.6$ &           $34.1 \pm 2.5$ &     $30.9 \pm 3.2$ &   $34.5 \pm 2.7$ \\ \addlinespace
\multirow{6}{*}{\textvert{\name-unseen}} &             \multirow{6}{*}{\xmark} &             32 &      $.84 \pm .07$ &          $24.4 \pm 2.1$ &   $32.1 \pm 2.7$ &         $.98 \pm .00$ &      $30.0 \pm 2.6$ &           $34.2 \pm 2.6$ &     $30.8 \pm 3.9$ &   $36.4 \pm 3.3$ \\
 &       &             16 &      $.83 \pm .07$ &          $24.2 \pm 2.1$ &   $32.7 \pm 3.1$ &         $.98 \pm .00$ &      $29.9 \pm 2.6$ &           $34.1 \pm 2.7$ &     $30.3 \pm 3.6$ &   $36.0 \pm 2.7$ \\
 &       &              8 &      $.82 \pm .08$ &          $23.8 \pm 2.0$ &   $32.6 \pm 3.2$ &         $.98 \pm .00$ &      $29.9 \pm 2.6$ &           $34.2 \pm 2.6$ &     $30.7 \pm 3.7$ &   $35.8 \pm 2.8$ \\
 &       &              4 &      $.81 \pm .08$ &          $23.3 \pm 1.9$ &   $32.2 \pm 3.2$ &         $.98 \pm .01$ &      $29.9 \pm 2.6$ &           $34.1 \pm 2.7$ &     $30.7 \pm 3.9$ &   $35.2 \pm 2.5$ \\
 &       &              2 &      $.78 \pm .09$ &          $22.9 \pm 2.0$ &   $32.1 \pm 2.4$ &         $.98 \pm .01$ &      $29.6 \pm 2.5$ &           $34.0 \pm 2.6$ &     $29.7 \pm 3.4$ &   $35.2 \pm 2.6$ \\
 &       &              1 &      $.74 \pm .11$ &          $22.1 \pm 2.0$ &   $31.9 \pm 2.9$ &         $.97 \pm .01$ &      $29.7 \pm 2.5$ &           $33.9 \pm 2.5$ &     $30.0 \pm 3.9$ &   $34.5 \pm 2.7$ \\
\bottomrule
\end{tabularx}
    \caption{Model scores (Dice for segmentation and skull-stripping, PSNR for other tasks) for each model and task as a function of the available subjects for training (U-Net) or context set (\name). Higher values are better. We average scores across all test subjects, eight modalities, and four segmentation classes (Cerebal cortex, Lateral ventricle, Thalamus, Hippocampus). Standard deviation across modalities and segmentation classes.}
    \label{tab:score_per_task}
\end{table*}

\clearpage

\section{Class names for Hammers Atlas dataset (experiment 3)} \label{supplement:hammers_atlas_class_names}
We provide label names and indices for the tissue classes in \cref{tab:hammers_atlas}, re-compiled from~\cite{Hammers2003,Gousias2008,Faillenot2017}.

\begin{table*}[h]
\centering
\begin{tabular}{lrl}
\toprule
Abbreviation & Class Index & Class Name \\
\midrule
Hip &    2      &   Hippocampus         \\
PAG &    10      &  Parahippocampal and ambient gyri          \\
STG &    12      &  Superior temporal gyrus          \\
MIG &    14      &  Middle and inferior temporal gyri          \\
FuG &    16      &  Lateral occipitotemporal gyrus (fusiform gyrus)          \\
Stm &    19      &  Brainstem          \\
Ins &    20      &  Insula          \\
PCG &    26      &  Gyrus cinguli, posterior part          \\
Tha  &    40      &  Thalamus          \\
CC &    44      &  Corpus callosum          \\
3V &    49      &  Third ventricle          \\
PrG &    50      &  Precentral gyrus          \\
PoG &    60      &  Postcentral gyrus          \\
ALG &    94      &  Anterior long gyrus          \\
\bottomrule
\end{tabular}
\caption{Hammers Atlas label abbreviations.}
\end{table*}

\clearpage
\section{Training dataset creation} \label{supplement:train_data}
We dynamically generate input image $x_t$, ground truth output $y_t$, and context set $\{(x_{t,j}, y_{t,j})\}_{j=1}^{N}$ from a collection of underlying datasets (\cref{tab:training-data}) during training.

In every training iteration, we first sample a task $t$ from $T_{\text{seen}}$. Next, one of the underlying datasets is selected to generate the sample $(x, y)$. Due to the makeup of the datasets, not every task can be performed on every dataset. For example, a dataset involving a single modality can not naturally be used to generate a modality transfer task. From the list of valid datasets, we sample the datasets for the input and context images independently, with a $1/3$rd chance of all context images coming from the same dataset as the input, $1/3$rd chance that context datasets are sampled at random from the valid datasets, and $1/3$rd chance that the context does not contain any subjects of the input dataset.

After the selection of task and dataset, we create the input and output images. This creation varies by task. We draw the subjects from each dataset at random, but exclude the input subject to re-occur as a context set member. For most tasks, we sample a subset of between one to three image modalities from the subject. For the segmentation task, we join a random subset of available segmentation classes into a binary target mask. For reconstruction and denoising tasks, noise and artifacts in the input images are simulated according to~\cite{Singh2022}. For the modality transfer task, we select a separate target modality. For the inpainting task, we create a random binary mask from Perlin noise mask these areas from the input image. For skull stripping, the target is a binary brain mask. For tasks other than segmentation and modality transfer, the modality of context images can vary from the input image.

\clearpage
\section{Inference cost and model size} \label{supplement:inference_cost}
We provide model parameter counts and inference costs. We use a Baseline U-net with 64 channels for experiments with limited data set sizes, and a U-Net with 256 channels for experiments on all data. For \name, we use the same model in all experiments, but the inference cost increases linearly with the size of the context set.

\begin{table}[h]
\centering
\caption{Model size and inference cost.}
\label{tab:my-table}
\begin{tabular}{lrr}
\toprule
Model                     & inference FLOP (g) & Parameters (m) \\ \midrule
Baseline, 64 channels     & 20.7               & 0.62           \\
Baseline, 256 channels    & 329.7              & 9.84           \\
\name, 1 ctx image   & 39.1               & 1.27           \\
\name, 32 ctx images & 610.5              & 1.27           \\
\bottomrule
\end{tabular}
\end{table}

\clearpage
\section{Task weights} \label{supplement:task-weights}
To speed up training, we use weighted sampling of tasks during training. Task weights are shown in \cref{tab:task-weight}. These values have been tuned experimentally. Tasks that converge fast and achieve high-quality results are given a lower weight. Tasks that take longer to converge or are given a higher weight.

\begin{table}[h]
\centering
\caption{Task weights during training.}
\label{tab:task-weight}
\begin{tabular}{lr}
\toprule
Task                                & Weight   \\ \midrule
Binary Segmentation                  & $2.0$        \\
Modality Transfer                    & $2.0$        \\
Superresolution                      & $1.0$          \\
Skull Stripping                      & $.5$           \\
Motioncorrection Reconstruction      & $.5$        \\
Denoising \& Bias correction         & $.5$          \\
k-space Undersampling Recon.         & $1.0$          \\
Inpainting                           & $1.0$\\
Simulated Modality Transfer          & $1.0$        \\
Masking                              & $.5$ \\  \bottomrule
\end{tabular}
\end{table}

\end{appendices}
\end{document}